\documentclass[aps,rmp,reprint,amsmath,amssymb,graphicx,longbibliography]{revtex4-1} 

\usepackage{bm,upgreek,bm,appendix,soul,color,graphicx}
\usepackage[normalem]{ulem}

\usepackage[T1]{fontenc} 

\usepackage{comment}
\usepackage{natbib}

\def\<{\langle}
\def\>{\rangle}

\newcommand{\bea}{\begin{eqnarray}}
\newcommand{\eea}{\end{eqnarray}}

\begin{document}

\raggedbottom 

\title{Room Temperature Superconductivity: \\
  the Roles of Theory and Materials Design}

\author{Warren E. Pickett}
\affiliation{Department of Physics and Astronomy,\\ 
University of California Davis, \\
Davis, California 95616}

\date{\today}

\begin{abstract}
For half a century after the discovery of superconductivity, materials exploration for better superconductors proceeded without knowledge of the underlying mechanism. The 1957 BCS theory cleared that up: the superconducting state occurs due to strong correlation in
the electronic system: pairing of electrons over the Fermi surface. 
Over the following half century 
 higher critical temperature T$_c$ was achieved only serendipitously as new materials were synthesized. Meanwhile the formal theory of phonon-coupled superconductivity at the material-dependent level became progressively more
highly developed: by 2000, given a known compound, its value 
of T$_c$, the corresponding superconducting gap function, and several other properties 
of the superconducting state became available independent of further experimental 
input. In this century, density functional theory based computational materials 
design has progressed to a predictive level -- new materials can be predicted from 
free energy functionals 
on the basis of various numerical algorithms. Taken together, these capabilities enable 
theoretical prediction of new superconductors, justified by application to superconductors 
ranging from very weak to quite strong coupling. Limitations of the current procedures 
are discussed briefly; most of them can be handled with additional procedures. 
Here we recount the process that has resulted in the three new highest 
temperature superconductors, 
with compressed structures predicted computationally and values of T$_c$ 
obtained numerically, 
that have been subsequently confirmed experimentally: the designed superconductors 
SH$_3$, LaH$_{10}$, YH$_9$. 
These hydrides have T$_c$ in the 200-260K range at megabar pressures; the experimental 
results and confirmations are discussed. While the small mass of hydrogen provides the 
anticipated strong coupling at high frequency, it is shown that it also enables 
identification of the atom-specific
contributions to coupling, in the manner that has been possible previously 
only for elemental superconductors. 
The challenge is posed: that progress in understanding of higher T$_c$ is limited by the
lack of understanding of screening of H displacements.
Ongoing activities are mentioned and current challenges are suggested, together with 
regularities that are observed in compressed hydrides that may be useful to  
guide further exploration.

\end{abstract}
\maketitle

\tableofcontents

\section{Prologue}
\begin{figure*}[tbp]
  \includegraphics[width=1.35\columnwidth]{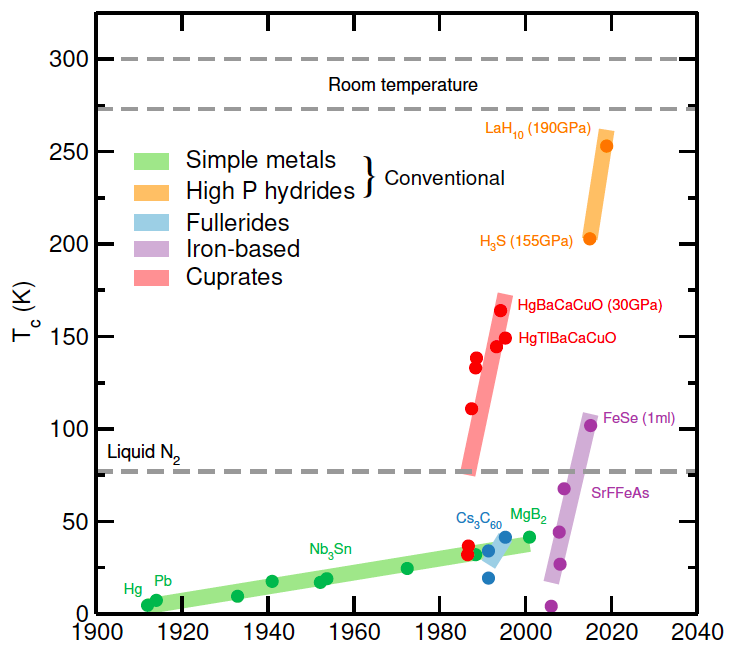} 
  \caption{A plot, from the discovery of superconductivity in Hg in 1911 and 
using linear scales, of the major advances in the maximum T$_c$ versus year. 
Advances within classes are denoted by thick linear areas, with color
delineating different classes. Certain temperature hurdles are denoted by 
dotted horizontal lines.
  High pressure hydrides lie at the upper right (yellow), with temperatures 
converging toward
room temperature. From Pickard {\it et al}. (2019) with permission.}
  \label{fig:maxTc}\end{figure*}

Room temperature superconductivity (RTS) has been one of the 
grand challenges of condensed matter physics since the BCS theory 
of pairing (see Sec. II.A) was proposed and its predictions verified. 
The remarkable electronic and magnetic properties of the 
superconducting state readily suggest revolutionary applications, 
both in the laboratory and in the public sector.
The slow progress in increase in the maximum critical temperature 
T$_c$ from 4K in 1911 to 23K in 1973 -- 3K/decade -- followed by a 
15 year plateau,  moved this grand challenge well into the background. 
The discovery of high T$_c$ cuprates (up to 134K, or 164K under 
pressure \cite{Gao1994}), involving magnetic interactions, 
introduced a new type 
of superconductivity (SC) and generated renewed excitement. 
After eight years of advancement of T$_c$, another plateau in 
cuprate superconducting T$_c$ has lasted for (so far) 28 years.
This recent advancement of the maximum T$_c$, revealing a  
breakthrough increase 
toward room temperature superconductivity that prompted this article, 
is shown in the upper right corner of Fig.~\ref{fig:maxTc}. 
After preliminary information,
in Secs. V and VI these advancements and some of their microscopic origins will
be discussed.

Much attention has centered on guidelines (`rules,' or `roadmaps') 
necessary for high T$_c$, thereby presumably pointing the way to yet higher T$_c$.
Such rules have been based primarily on known superconductors, and have subsequently been set aside 
as entirely new classes of superconductors were discovered, 
almost entirely by serendipity. In the 1960s empirical trends led to ``Matthias's Rules.'' These stated that for high T$_c$ one should search for 
(1) cubic materials, 
(2) $d$ electrons at the Fermi level, and 
(3) specific electron/atom ratios. This latter rule was soon understood to mean a high density of electron states at the Fermi level ($\varepsilon_F$=0)  N(0), {\it i.e.} a high density of superconducting pairs.
Bernd Matthias' group had been the leader in the discovery of new superconductors in the 1950s and early 1960s, 
after which the search extended worldwide. 

Though unwritten, additional rules were advertised: (4) stay away from oxygen, which produces unpredictable behavior including insulation, and (5) stay away from magnetism, which at the time competed too strongly with superconducting pairing. Matthias had an additional personal rule: (6) stay away from theorists, they are no help. (Matthias did write \cite{Matthias1972}, without elaboration ``I never realized how many of my friends are theorists.'') This last rule was based on his observation that knowing the theory of SC, at least in a broad model BCS way, had been useless in helping discovery; what worked (but ploddingly so) was just to ``follow the simple roadmap.'' However, the rules did not produce high T$_c$ materials. It continued to be true through cuprate days -- after the initial high T$_c$ -- that theory played no part in the advance of the maximum T$_c$ from 30K to 134K in cuprates. 

Over the intervening decades theory and numerical implementation 
have advanced. While overt activity toward higher BCS superconductivity waned, intellectual interest in the goal of RTS persisted, with evidence given by various international workshops indicating continued emphasis on room temperature 
explorations.\footnote{\label{Workshops}
This list provides six of the meetings.\\
$\bullet$1994 The Road to Room Temperature Superconductivity,
Bodega Bay, California, 1992.\\
$\bullet$2005 The Possibility of Room Temperature Superconductivity. Notre Dame, June 10-11, 2005.  https://www3.nd.edu/~its/rts/index.html\\
$\bullet$2014 (Toward) Room Temperature Superconductivity, Leiden, The Netherlands, 2014..\\
$\bullet$2016 SUPERHYDRIDES Towards Room Temperature Superconductivity: Hydrides and More, Rome, Italy, 2016.\\
$\bullet$2017  Towards Room Temperature Superconductivity: Superhydrides and More. Chapman College, California, 2017.\\
$\bullet$2022 Challenges in Designing Room Temperature Superconductors.
L'Aquila, Italy, 2022.}
With the 2014-2015 discovery of extreme high T$_c$ SC in compressed metal hydrides under pressure discussed in this paper, the roles of experiment and theory evolved and were reversed. Theory has assumed a prominent role as predictor beyond the maximum known T$_c$ (T$_c^{max}$) for phonon-coupled SCs as they jumped from 40K to 200K, then rapidly marched toward room temperature. 


After recounting the sequence of necessary theoretical advances in Sec. II, 
Sec. III gives a brief indication of the  computational innovation 
and implementations that were 
required to design real, heretofore unknown, materials.  An overview of 
the rising research area in this century of crystal structure prediction is 
given in Sec. IV. Section V provides a concise description of the first three 
revolutionary discoveries of critical temperatures in the 200-260K range.
So far these (and a few others) all require megabar pressures. The discussion in 
Sec. VI reveals how the light H atom restores the sort of analysis that was developed 
for elemental metals, which leads to demonstration that H dominates the metal 
component in promoting very high T$_c$; in fact, the contribution of the metal 
atom provides confusion by contributing in opposing ways to the properties that
promote high T$_c$. This leads to Sec. VI, which gives an 
overview of near-term challenges and opportunities that have been identified in 
the very high T$_c$ arena. These are provided in terms of several regularities of high T$_c$ 
compressed hydrides that may guide the next level of searches, as compiled in 
Sec. VII. Section VIII provides a brief Epilogue.

\section{Theoretical Developments through the Decades}
The impetus for this stunning breakthrough has been a sequence of advances in 
theory, numerical implementation, and computational design together with mastery  
of high pressure techniques. Since this advance in computational theory has 
extended over six decades ({\it i.e.} three generations of physicists), we 
provide in this section an overview of the fertile path of breakthroughs in theory 
and numerical implementation that have enabled current capabilities and their 
paradigm-revising results.

\subsection{BCS Theory}
In 1957 Bardeen, Cooper, and Schrieffer published their 30-page {\it magnun opus} ``Theory of Superconductivity,'' \cite{BCS} that includes around 275 numbered equations. This theory introduced a correlated manybody wavefunction based on Fermi surface pairing of electrons demonstrated a year earlier by Cooper. This wavefunction described a thermodynamic condensate of correlated pairs below a critical temperature T$_c$, assuming an attractive effective pairing potential arising from exchange of phonons and screened electron-electron scattering. The primary result of their theory was the T$_c$ equation (units $\hbar$=1, $k_B=1$ will be used in this article)
\bea
T_c^{BCS} = 1.14~\Omega~e^{-1/\lambda^*}
\label{eqn:BCS}
\eea
where $\Omega$ is the characteristic phonon frequency.  
$\lambda^*=N(0)V = \lambda - \mu^*$ in terms of terminology developed later (see below). 
The Fermi level density of states $N(0)$ is a measure of the density of electrons 
interacting through the phenomenological coupling $V$, which BCS noted is the 
${\it net~interaction}$ of phonon attraction plus Coulomb repulsion.  

The electron-phonon coupling strength $\lambda$ is 
the measure of the attractive pair coupling strength by exchange of phonons. Out of 
interest for the following content of this article we note that the strong coupling 
limit $\lambda^*\rightarrow\infty$ leads in this expression to a maximum SC temperature 
of 1.14~$\Omega$, which can be well above room temperature. The theory is however 
only valid for weak coupling, leaving open the question of high T$_c$ and the 
large $\lambda$ regime. ``High T$_c$'' would only be confronted three decades later 
with the discovery of cuprate superconductors, for which there is no predictive 
theory and hereafter will not be considered.

\begin{figure}[tbp]
 \includegraphics[width=1.00\columnwidth]{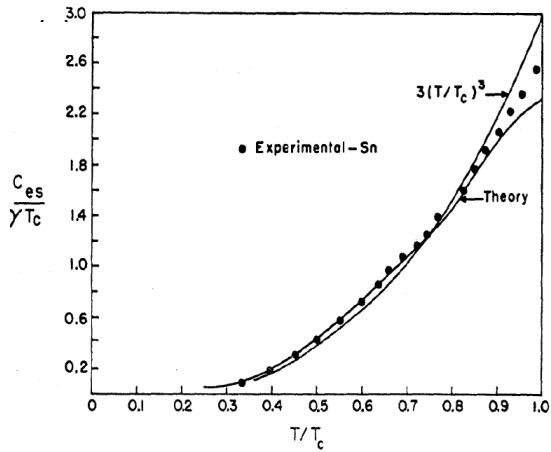}
\vskip -5mm
  \caption{Figure from the BCS paper [Bardeen {\it et al.} (1957)] 
paper illustrating the predicted temperature 
dependence of the ratio of the specific heat in the superconducting state to that in
the normal state value (equal to 1.0 on the abscissa). The behavior is exponentially 
small at low temperature due to the energy gap, in excellent agreement with experimental
data. }
  \label{fig:BCS}
\end{figure}

The second property of note, provided by BCS theory and its extensions, is the superconducting gap [$\Delta(\omega,T)$] equation that gives T$_c$ when linearized ($\Delta\rightarrow 0)$ and at finite temperature provides the frequency $\omega$ and temperature $T$ dependence of the energy gap that is responsible for many of the fascinating properties of superconductors. The novelty is that a perfect conductor can have a gap for excitations; this gap leads to a vanishing magnetic susceptibility in its interior; it is a basic player in zero resistivity; it severely alters thermodynamic, optical, and transport properties. 

The effect on the electronic specific heat, shown in Fig.~\ref{fig:BCS}, is to make it vanish exponentially as T$\rightarrow$0 rather than linearly to zero as for the normal metal. The theory was soon generalized by Gor'kov, using thermodynamic field theory, to derive \cite{Gorkov} the much used 1950 phenomenological free energy theory of Ginzburg and Landau. \cite{ginzburg} This extension of theory also provided a direction for pursuing necessary developments, especially a microscopic understanding of total electron-electron interactions in metals, a path that is discussed below. 

\subsection{Migdal-Eliashberg Theory}
The microscopic Migdal+Eliashberg theory of 1960 (henceforward called Eliashberg theory)  of electron-phonon (EP) coupled superconductivity, weak or strong, followed very quickly after the BCS theory (1957) based on a model pairing  Hamiltonian and a variational treatment. Recall that the full, exact (non-relativistic) Hamiltonian ${\cal H}$ for a system of ions \{$\vec R$\} and electrons \{$\vec r$\} is given by 
\bea
{\cal H}&=&[{\cal T}^{el}(\{\vec r\}) + {\cal V}^{el-el}(\{\vec r\})] \\ \nonumber
&+&  {\cal V}^{el-ion}(\{\vec r\},\{\vec R\}) \\ \nonumber
&+&  {\cal T}^{ion}(\{\vec R\})+
     {\cal V}^{ion-ion}(\{\vec R\})
\label{eqn:hamiltonian}
\eea
in terms of the various kinetic ${\cal T}$ and potential ${\cal V}$ energies. The middle term contains the (bare) electron-ion interaction. With roughly 10$^{20}$ dynamical coordinates per mm$^3$, this all-encompassing operator is the fundamental, intractable feature of materials theory that requires innovative approaches and reliable approximations.

Migdal formalized the observation \cite{ME1} that the great differences in energy scales (or velocities, or masses) of electrons and ions leads to negligible contributions beyond second order perturbation theory in EP scattering.  Eliashberg made the formidable step of placing the pairing theory within the newly developed many-body theory (thermal Green's function) that made it applicable to the superconducting state, \cite{ME2,ME3} leaving generalizations to materials with crystal structure and lattice effects as a later step (see below). This Eliashberg formalism provided the fundamental equations for calculating the gap function (superconducting order parameter), given the Eliashberg spectral function $\alpha^2F(\omega)$ defined in Sec. II.E that provides the essential input for the pairing process. Calculation of this function necessitated several theoretical and algorithmic advances.


\subsection{Density Functional Theory (DFT)}
Hohenberg, Kohn, and Sham followed immediately (1965-1967) with density functional theory, \cite{DFT1,DFT2} which makes the full crystalline Hamiltonian in Eq.~2   
treatable (for static nuclei) for many properties of materials for electrons in their ground state, given a reasonable approximation for many body effects (through an exchange and correlation functional). (The many following extensions of DFT to other properties are not relevant here, except as noted.) 

An extremely important aspect of DFT is that it provides a highly reliable one-electron (`mean field') set of Kohn-Sham one-electron band energies and wavefunctions \cite{DFT2} for use in themselves, and also for applications in treating dynamic behavior that is averaged over in DFT.  This DFT breakthrough accelerated activity in band theory, already in progress since the late 1930s based on more phenomenological grounds. The late 1970s saw the accomplishment of achieving self-consistent electronic charge densities, bands, and wavefunctions. Many applications of (electronic) ground state energies were explored in the 1980s and following. 

Calculation of energies for any configuration of atoms enabled the evaluation of interatomic force constants between displaced atoms, and thereby phonon spectra. Initially this was accomplished  
for each phonon independently, but capabilities were rapidly extended, especially by applying density functional perturbation theory and Wannier function techniques. \cite{Wannier,latticewannier,cockayne} The outcome was true harmonic phonon frequencies through formalism making use of infinitesimal atomic displacements. The change in the electronic potential due to an atomic displacement is also the root factor in EP coupling, and calculation of EP matrix elements was achieved only around 2000.

\subsection{Extension of Eliashberg Theory to Real Materials}
Also in the mid-1960s, the challenge of addressing real superconductors versus simplified models, a prerequisite for materials design and discovery, was accomplished by Scalapino {\it et al.} in 1964. \cite{Scalapino1,Scalapino2} Starting from the full Hamiltonian of a solid (Eq.~\ref{eqn:hamiltonian}) they derived a remarkably complete formalism for the superconducting Green's function and the full frequency and temperature dependence of the complex gap function.   Their formalism awaited a viable description of the underlying electronic energy bands and phonon frequencies, and their coupling. The DFT capabilities discussed in Sec.~II.C would provide the underlying electronic bands and wavefunctions, phonon dispersion curves and polarizations, and EP matrix elements.

The validity of their formalism, which underlies today's numerical implementation, relies on a few approximations. One is Migdal's theorem mentioned above: the vast differences in masses of electrons and ions (more precisely, differences in the frequencies of their dynamic responses) specifies that second order perturbation theory is sufficient for the electron and phonon self-energies. Secondly, electron-electron Coulomb repulsion effects leave the metal as a conventional Fermi liquid; specifically, the possibility of magnetic order and magnetic fluctuations is not included at the level we discuss here. 

Thirdly, the symmetry that is broken at the superconducting phase transition is so-called gauge symmetry; in picturesque language, two electrons can form a Cooper pair and disappear into the condensate, or the inverse process can happen. In this language, electron number conservation is broken. [C. N. Yang liked to emphasize that electrons do not actually disappear, and that the fundamental signature of the superconducting state is appearance of long-range order in the two-particle density matrix.] Finally, for EP-coupled superconductors it is nearly always the case that the superconducting order parameter is proportional to the gap function, {\it i.e.} a complex scalar and not a vector or tensor quantity, and other symmetries are not broken at T$_c$. The resulting theory has passed numerous tests as being accurate for EP-coupled superconductors.

\subsection{Analysis of T$_c$; discussion of a maximum}
 \subsubsection{McMillan's analysis}
In 1967 McMillan advanced the analysis of the origins of T$_c$ substantially.
 \cite{McMillan}
He defined the electron-phonon spectral function $\alpha^2F$, also called the Eliashberg function, 
in terms of the various quantities appearing in the Scalapino {\it et al.} expressions:
\bea
\alpha^2F(\omega)=N_{\uparrow}(0)
 \frac {\sum_{kk'}|M_{kk'}|^2 \delta(\omega-\omega_{k-k'})
   \delta(\varepsilon_k)\delta(\varepsilon_{k'}) }
  { \sum_{kk'}\delta(\varepsilon_k)\delta(\varepsilon_{k'}), }
  \label{eqn:a2f}
\eea
where the Fermi energy $\varepsilon_F$=$0$, and $k$-$k'$ is the wavevector of the phonon scattering an electron from state $k$ to $k'$, each confined to the Fermi surface. Necessary sums over bands and phonon branches are not displayed explicitly. The sums that are shown are each over the three dimensional Brillouin zone. 
The frequency $\delta$-function
provides the frequency resolution of coupling.  For clarity, the density of states factor is for a single spin, designated by the arrow on $N_{\uparrow}(0)$. 

The EP coupling strength $\lambda$, and the two frequency moments $\omega_{log}$ 
and $\omega_2$ used prominently
by Allen and Dynes (see the next subsection) are given in terms of
moments of 2$\alpha^2F(\omega)/\omega$ by
\begin{eqnarray}
\lambda    &=&
               \int d\omega          \frac{2\alpha^2F(\omega)}{\omega}, \\
\omega_{2}^{2} &=&
         \int d\omega \omega^2 \frac{2}{\lambda}\frac{\alpha^2F(\omega)}{\omega}
        \equiv \int d\omega \omega^2 g(\omega), \\
\omega_{log}&=&exp\left[\int d\omega \log\omega g(\omega)\right].
\end{eqnarray}
Here $g(\omega)$, defined by the second line, is the normalized shape function
of $\alpha^2F{\omega}/\omega.$
Aside from normalization, these are all different moments of 2$\alpha^2F(\omega)/\omega$. 
The calculation of $\alpha^2F$, necessary for the computational program to advance,
is discussed further in Sec. III.B.

\begin{figure}[tbp]
 \includegraphics[width=1.00\columnwidth]{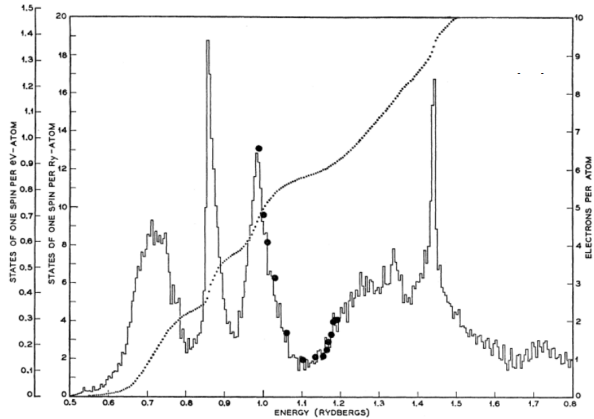}
\vskip -5mm
  \caption{McMillan's plot, for $5d$ transition metal alloys, 
of the experimental values 
(dots) of the band structure Fermi level densities of states, for which the mass 
enhancement 1+$\lambda$ of the specific heat constant$\gamma$ has been removed, compared 
to the band theory calculation of tungsten by L. F. Matthiess. The agreement is 
remarkable. Figure taken from McMillan (1967).}
  \label{fig:McM_NbDOS}
\end{figure}

For elemental metals,
McMillan obtained initially two fundamental expressions from $\alpha^2F$. First is that
its first frequency moment is {\it independent of vibrational properties} aside 
from an overall inverse mass factor:
\bea
\int \omega \alpha^2F(\omega)d\omega=\frac{N_{\uparrow}(0){\cal I}^2}{2M},
\eea
where ${\cal I}^2$ is the average atomic scattering strength by phonons from the Fermi
surface to the Fermi surface:
\bea
I_{k,k'}  &=& <k|\frac{dV(r;\{R\})}{dR_j}|k'>,\nonumber \\
M_{k,k'} &=&\sqrt{\frac{\hbar}{2M\omega_{k-k'}}}I_{k,k'}, \nonumber \\
{\cal I}^2      &=& <|I_{k,k'}|^2>_{FS}.
\label{eqn:matrixelement}
\eea
Here $V(r;\{R\})$ is the total electronic potential which depends on all ionic 
coordinates $\{R\}$, $R_j$ is the coordinate of the displaced atom under 
consideration, and $I_{k,k'}$ is the electron-ion matrix element between electron states
$k$ and $k'$. $M_{k,k'}$ is the electron-phonon matrix element that contains 
the atom mass M and phonon frequency $\omega_{k-k'}$, and  
the outside brackets indicate that an average
is taken over all $k$ and $k'$ both on the Fermi surface to obtain ${\cal I}^2$. 
This first-order
change in potential $dV/dR$ has a long history \cite{Pickett1979} 
and much progress in numerical
evaluation, without much progress in understanding. This integral expression 
holds only for elements with mass $M$, a restriction we manage to relax for hydrides in
Sec. VII.B.

McMillan's second observation was that (again for elemental metals) $\lambda$ can be
expressed simply in terms of physical quantities
\bea
\lambda = \frac{N_{\uparrow}(0){\cal I}^2} {M\omega_2^2} \equiv \frac{\eta}{\kappa},
\label{eqn:lambda}
\eea
The numerator $\eta=N_{\uparrow}(0){\cal I}^2$ is the McMillan-Hopfield factor that reflects an ``electronic stiffness,'' divided by  $\kappa=M\omega_2^2$ which is the standard harmonic oscillator form of lattice stiffness (mean force constant), with $\omega_2$ representing the entire spectrum of frequencies. All are normal state properties.
This relation indicates that increasing the frequency scale $\omega_2$, which helps T$_c$, will decrease $\lambda$ according to its square, hurting T$_c$, with net outcome yet to be understood.

An aside, for now. The simple expression Eq.~\ref{eqn:lambda} does not apply to compounds, 
as the quantities $N_{\uparrow}(0)$ and $\omega_2$ have contributions from all atoms. 
and ${\cal I}^2$ and $M$ are distinct atomic quantities.
It will be seen in Sec. VI.B that this simple `elemental' expression for $\lambda$
can be extended to compressed metal hydrides to obtain these quantities separately 
for H and the metal atom, thereby allowing deeper analysis into the origins of 
high T$_c$.

Most prominently McMillan provided, with justification for chosen algebraic forms,
 a seminal analysis \cite{McMillan} from Eliashberg theory of T$_c$ and its 
dependence on $\alpha^2F(\omega)$, presenting his iconic ``McMillan equation'' expressing T$_c$ via two materials quantities, the phonon (Debye) frequency $\theta$ and $\lambda$. A phenomenological retarded Coulomb repulsion $\mu^*=0.13$ was included in the Eliashberg equation that was solved to relate T$_c$ to $\alpha^2F$. Solutions for various $\alpha^2F$ functions were fit to a generalization of the BCS equation  of the form
\bea
T_c^{McM} &=& \frac {\theta}{1.45} 
  exp\left [ -1.04\frac{1+\lambda} {\lambda-\mu^* - 0.62\lambda\mu^*}\right] \nonumber \\
  &\equiv& \frac{\theta}{1.45} 
     exp \left(-\frac{1+\lambda} {\lambda-\mu_{eff}^*}\right )^{1.04}
    \label{eqn:McM}
\eea
where the second expression emphasizes the generalization of the 
BCS relation. The constants provided best fits to the computational data:
the `exponent' 1.04 seems clearly to be a parameter giving best fit
without any physical significance. 
Note that the McMillan equation is not appropriate for very weak coupling; 
when the denominator of the argument of the $exp$ function becomes negative, 
the result is non-sensical. What the McMillan equation does is to replace
the very involved functional dependence T$_c$($\alpha^2F,\mu^*$) by a
simple equation for T$_c$ involving two quantities from $\alpha^2F$ plus
$\mu^*$.

The ``effective Coulomb repulsion'' $\mu_{eff}^*=\mu^*(1+0.62\lambda)$ is 
defined here (without physical justification) to emphasize the similarity 
to the BCS equation (1). The $1+\lambda$ factor in the numerator is often 
assigned to the strong-coupling electron mass renormalization by EP coupling. 
The mass enhancement is well known to increase the density of active states at the Fermi level by 1+$\lambda$, but in this expression the effect in the numerator is to decrease T$_c$, compensated somewhat by the $\lambda\mu^*$ term in the denominator. Flores-Livas {\it et al.} have reproduced the model that gives a similar result, for $\mu^*(\lambda=0)$ and a different expression for the prefactor. \cite{FloresPerspective2020}

In McMillan's time the experimental Debye frequency $\theta$ for elemental metals
was the most accessible measure of the frequency spectrum, however McMillan 
recognized that using moments from $\alpha^2F$ (such as the second moment 
$\omega_2$) would be more representative. 

Only values of $\lambda$ to around unity were included in McMillan's fit; 
extrapolation of T$_c^{McM}$ shows that it saturates for large $\lambda$, 
however the equation is justified only for moderately strongly coupled 
superconductors.
The limited range  of input has been missed or forgotten by many readers, and 
until the discoveries discussed in the next section one could still find 
statements in the literature and in presentations that ``theory shows 
that the maximum T$_c$ is limited, perhaps to around 40 K.'' This point 
will be clarified below.

 \subsubsection{Bergmann and Rainer's functional derivative}
In 1973 Bergmann and Rainer \cite{bergmann} raised a straightforward question:
to increase T$_c$, one needs to understand the effect on T$_c$ due to an 
added increment of coupling $\Delta\alpha^2F(\omega)$ at frequency $\omega$. 
It seemed clear from the BCS equation that adding coupling at high frequency, 
seemingly increasing both $\omega$ (or $\Omega$, or $\theta$, depending on 
treatment of the phonon system) and $\lambda$ (adding an increment), 
would raise T$_c$. 
Study by Allen \cite{Allen1973a,Allen1973b} soon after McMillan gave suggestions 
that, due to this complex interdependence of materials properties, including 
$N_{\uparrow}(0)$ and ${\cal I}^2$, coupling at low frequency might even be harmful for T$_c$. 
We will see in Sec. VI.B that compressed hydrides provide an unexpected field 
of study of this unsettled question.

Bergmann and Rainer addressed the question by calculating the differential 
quantity in the kernel of
\bea
\Delta T_c=\int d\omega\frac {\delta T_c[\alpha^2F,\mu^*]} {\delta\alpha^2F(\omega)}
  \Delta \alpha^2F(\omega),
\eea
this being the functional derivative of T$_c$ with respect to $\alpha^2F(\omega)$:
given an increase in coupling by an increment $\Delta\alpha^2F$ at frequency $\omega$, 
what is the change $\Delta T_c$?

\begin{figure}[tbp]
 \includegraphics[width=1.10\columnwidth]{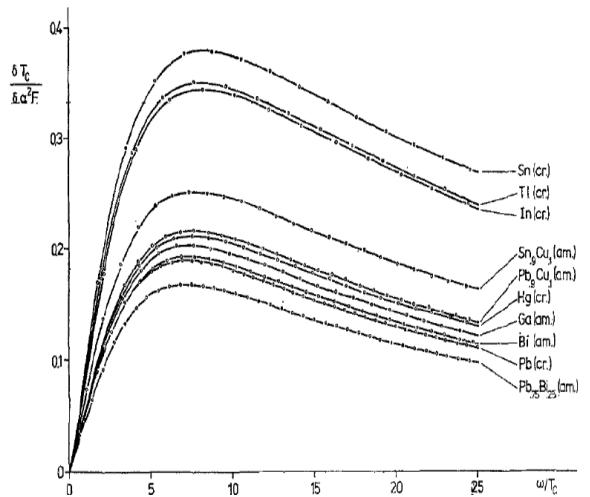}
\vskip -5mm
  \caption{
Bergmann and Rainer's plots of 
  $\delta T_c/\delta \alpha^2F(\omega$) versus 
the ratio $\omega/T_c$ for a number of superconductors (as labeled). 
Its value peaks near $\omega=2\pi T_c$ and decreases slowly thereafter, indicating 
little importance of low energy phonons, great importance for higher energies.
Figure taken from Bergmann and Rainer (1973).} 
  \label{fig:Bergmann}
\end{figure}

They established that this functional derivative is smooth ($\alpha^2F$ is not) and 
non-negative, thus every phonon helps, as anticipated. At least for conventional 
shapes and strengths of $\alpha^2F$ it has a broad peak around 
  $\omega_{BR}\sim$ 2$\pi$T$_c$ followed by a slow decrease at high frequency, 
shown in Fig.~\ref{fig:Bergmann}. Conversely, below $\omega_{BR}$ the usefulness 
of coupling decreases linearly to zero. Low frequencies can and do contribute 
strongly to $\lambda$: viz. the corresponding functional derivative is
$\delta\lambda/\delta\alpha^2F(\omega)=2/\omega$. 

This result, viz. that coupling at high frequencies is most important, does not 
quite resolve all issues. While adding some differential coupling 
$\Delta \alpha^2F(\omega)$ does affect both $\lambda$ and $\omega$ calculated 
from $\alpha^2F$, within the larger picture it also will affect the EP coupling 
that causes changes in $\alpha^2F$ at frequencies besides $\omega$. This effect 
is not settled by simply including a single (extrinsic) increment in $\alpha^2F$. 
It will be seen in Sec. VI.B that hydrides provide counterexamples to the simple
interpretation offered by the functional derivative of Bergmann and Rainer.

 \subsubsection{Allen and Dynes' reanalysis}
Building on the work of McMillan, Allen and Dynes in 1975 presented a 
re-analysis \cite{AllDyn} of T$_c$, using more than two hundred solutions of the 
Eliashberg equation for $\lambda$ up  to 10 and $\mu^*$ up to 0.20, including 
experimental determinations of $\alpha^2F$ (by numerical inversions of tunneling 
data) with associated measured T$_c$. Allen and Dynes chose to generalize the 
McMillan equation. Most obvious is that the frequency prefactor is updated 
from $\omega_2$ (altered earlier from McMillan's $\theta$) to the logarithmic 
moment $\omega_{log}$ with an adjusted constant, as the latter produced a 
more consistent fit to their data. 

\begin{figure*}[tbp]
 \includegraphics[width=2.10\columnwidth]{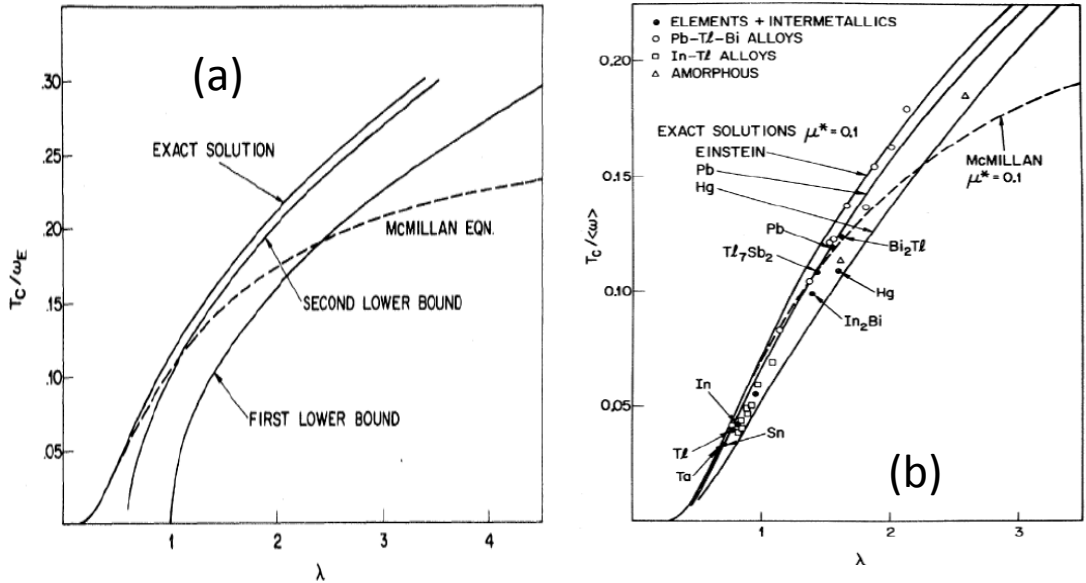}
\vskip -20mm
  \caption{
(a) Allen and Dynes' plots of the ratio $T_c/<\omega>$ versus $\lambda$ for
the McMillan equation, and for truncated and exact numerical results from the
Eliashberg equation. 
(b) Similar plots (also from Allen and Dynes (1975)) using data
for several known strong coupled superconductors. Also shown is the exact
solution for an Einstein model with frequency $<\omega>=\Omega$, with the deviation from
the McMillan equation beginning at $\lambda\sim 1$. Here $<\omega>$ is the
conventional first moment of $\alpha^2F(\omega)/\omega$. Figures are
from Allen and Dynes (1975) with permission. }
  \label{fig:AllDyn}
\end{figure*}

More importantly, they observed that corrections were needed for both the strength of coupling -- they considered much larger values of $\lambda$ than previously studied -- 
and for unusual shapes of the $\alpha^2F/\omega$ spectrum. For example, the textbook shape of the Nb phonon spectrum is much different from that of (say) PdH, which consists of low frequency Pd acoustic modes separated from high frequency H optic modes. We return to the implications of this two-peak shape in Sec. VI.B. 

The shape dependence was framed by Allen and Dynes in terms of the ratio
$\omega_2/\omega_{log}$, which is always greater than unity. In their fit to extensive data, the argument of the exponent was not changed, rather each change (strong coupling and shape) was incorporated into its own prefactor $f_1$ and $f_2$ respectively. The Allen-Dynes equation can be written, considering the McMillan form given above,
\bea
T_c^{AD}&=& \frac{\omega_{log}} {1.20} \frac{1.45}{\theta}
 f_1(\lambda,\mu^*)
   f_2(\omega_{log},\omega_2) T_c^{McM}(\lambda,\mu^*) \nonumber \\
 &=& \frac{\omega_{log}} {1.20}     f_1 f_2 
   exp \left(-\frac{1+\lambda} {\lambda-\mu_{eff}^*}\right )^{1.04} \nonumber\\ 
 &=& \frac{\omega_{log}} {1.20}     f_1 f_2
   exp\left [-1.04\frac {1+\lambda} 
        {\lambda-\mu^* - 0.62\lambda\mu^*}\right ],
        \label{eqn:AD}
\eea
the latter form being the more common one and the one presented by Allen and Dynes. 
The simple expressions for $f_1$ and $f_2$ can be found in the original 
paper. \cite{AllDyn}  
Note that $\theta$ cancels in the first expression, and is replaced by the 
logarithmic frequency moment, which Allen and Dynes found to be the most useful 
moment for the prefactor of T$_c$, at least for $\lambda$ up to two. 

Due to the $f_1$ and $f_2$ factors, T$_c^{AD}$ is no longer exponential 
in the argument involving $\lambda$ except in the weak-coupling region 
where $\lambda$ begins to approach $\mu^*$. In this regime 
a solution of the Eliashberg
equation, as well as T$_c^{AD}$, is highly sensitive to each quantity. 
For this reason, their equation is not a meaningful fit to any data in this 
weak coupling region. 
For the high T$_c$ compressed hydrides discussed in Sec. V, each prefactor can 
approach a 10-15\% enhancement of T$_c$ (or increasingly more as $\lambda$ 
is increased still further). There is no reason not to use the Allen-Dynes equation except to observe the relative importance of the $f_1$ and $f_2$ corrections, and $\omega_{log}$ instead of an alternative frequency.

A crucial finding of Allen and Dynes was that in the large $\lambda$ regime, where 
analytic results could be extracted, the asymptotic behavior is
\bea
T_c \rightarrow 0.18 \sqrt{\lambda\omega_2^2}=0.18\omega_2\sqrt{\lambda}=0.18\sqrt{\eta/M}
\label{eqn:limit}
\eea
for $\mu^*=0.10$ (the prefactor depends somewhat on $\mu^*$). 
Here $\lambda=\eta/(M\omega_2^2)$ for elements has been used to display different 
viewpoints. Note that the asymptotic limit involves only the same frequency independent 
constant identified by McMillan.  A primary implication is that Eliashberg theory poses 
{\it no limit on T$_c$}. Figure~\ref{fig:AllDyn}(a) shows some of the numerical 
data that establishes the ever-increasing (not plateauing) of T$_c(\lambda)$, and panel (b)
of this figure  provides numerical data using experimental shapes of $\alpha^2F$.

An informative observation is that, as far as the limiting value goes, neither $\lambda$ nor $\omega_2$ are separately relevant; physically, increasing $\lambda$ decreases $\omega_2$ and there are numerous examples of this. 
It should be recognized that performing this limit assumes nothing else happens except that $\lambda$ increases. 
This procedure is non-physical; making changes that increase $\lambda$ in turn decrease frequencies due to the increased coupling from electronic states at higher energy. Increasing $\lambda$ often creates lattice instabilities, so the more strongly coupled material cannot exist. 
On the positive side, nothing prohibits finding superconductors with 
increasingly larger $\lambda$ 
with appropriate frequencies and higher T$_c$.  Further analysis of T$_c$ can be 
found in extended descriptions by Allen \cite{Allen1979,Allen1980} and 
Allen and Mitrovi\'c. \cite{Mitrovic}

\subsection{Solidifying the formal theory}

\begin{figure}[tbp]
  \hskip -4mm
  \includegraphics[width=0.85\columnwidth]{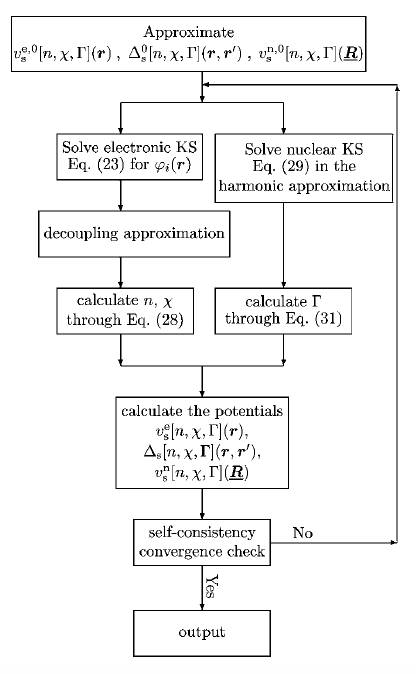} 
 \vskip -3mm
  \caption{Flow chart for the {\it ab initio} (entirely empirical-information free)
calculation for superconducting density functional theory, 
describing electrons and nuclei in
thermal equilibrium in a superconducting state.
Reproduced with permission from \onlinecite{SCDFT1}, 
from which the definitions of various items and the full
formalism can be obtained.}
  \label{fig:scDFT1}
\end{figure}

\subsubsection{General remarks}
Eliashberg theory providing T$_c$, the frequency and temperature dependent gap
$\Delta(\omega,T)$, and more,  as described above and as commonly applied, 
is based on the common but somewhat {\it ad hoc} choices of 
(i) using the DFT mean-field 
eigenvalues and eigenfunctions as including the essential electronic 
interaction effects in a static lattice (this choice has no rivals), 
(ii) from that, the self-consistently determined phonon spectrum including 
static electronic screening, as obtained from DFT, and 
(iii) the simple bubble diagrams giving the EP coupling contribution to
the electron and phonon self-energies (always consistent with Migdal's theorem).
The Hamiltonian then consists of band (bare) electrons (in the static lattice),
bare harmonic phonons oscillating in the frozen electron density {which are
non-physical and ill-defined), and (iv) the
bare electron -- bare phonon coupling due to first-order change in the lattice
potential due to atomic motion. This step-by-step {\it ansatz}, though seemingly 
first principles, leaves questions about the
full self-consistent treatment of the Coulomb interaction.
 
A general formulation of the electron-phonon problem based on the full
crystal Hamiltonian, provided by 
Allen, Cohen, and Penn, \cite{ACP_PRB1988} 
reveals that the combined electron+lattice polarization (and resulting
dynamic dielectric screening function) is the central quantity to be addressed 
by many-body theory. While treating the polarizations separately has become
intuitive and works well for conventional metals, 
a full treatment might reveal novel processes and possibly new phases of 
matter in complex materials.

A rigorous underlying formalism for a complex problem often guides progress even when 
rigor must be relaxed. DFT provides an excellent example. DFT gives a rigorous foundation 
for treating the full crystal Hamiltonian in Eq.~\ref{eqn:hamiltonian}, with the 
restriction to static nuclei. Beyond the ground state energy and magnetic properties, 
it is established that the Kohn-Sham band structures are reasonable to excellent 
representations of single particle excitations in conventional (Fermi liquid)
metals, as soon as a sufficiently sophisticated 
exchange-correlation functional is incorporated. This characterization is especially 
relevant near the Fermi surface of metals, where necessary corrections can be 
incorporated when desired.
 
\subsubsection{Superconducting density functional theory}
Such a formal underpinning for superconductors was devised by Gross and a sequence 
of collaborators. For this article we emphasize the density functional theory for 
superconductors (SCDFT) formulated for material specific studies, which was achieving 
implementation by 2005.\cite{SCDFT2}.  The formulation includes a number of innovations, 
with the basic ones being the extension of ground state DFT to thermodynamic DFT, 
followed by generalization to description of quantum nuclear degrees of freedom, then 
incorporating an allowance for a superconducting order parameter.

\begin{figure}[tbp]
  \hskip -4mm
  \includegraphics[width=0.85\columnwidth]{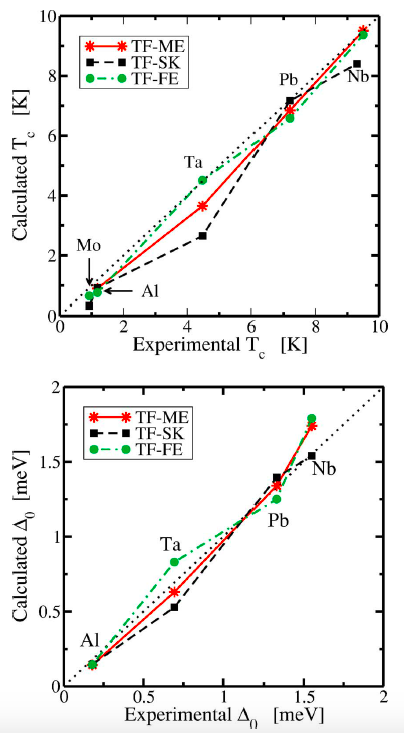} 
 \vskip -3mm
  \caption{Top panel: T$_c$ calculated within SCDFT plotted versus the experimental value.
The dotted line along the diagonal indicates perfect agreement. The different 
piecewise linear lines
indicate the use of three different dielectric functions for the Coulomb repulsion.
Results for the TF-FE choice are almost indistinguishable from the experimental value.
  Bottom panel: the analogous plot for the zero temperature static energy gap
(calculations done at 0.01 K), which can be measured in several experiments.
All show extremely encouraging agreement; the TF-SK values are somewhat better 
than TF-FE values compared
to experiment. No information of the Mo superconducting  gap 
(which is very small) is available.
Reproduced from \onlinecite{SCDFT2} with permission.
}
  \label{fig:scDFT2}
\end{figure}

Accounting for superconductivity involves a functional $F[n,\chi,\Gamma]$ of the 
electron density $n(\vec r)$, the superconducting order parameter 
$\chi(\vec{r},\vec{r}')$, and the diagonal of the nuclear density matrix 
$\Gamma(\{R\})$. The resulting three Euler-Lagrange equations involve one for 
the nuclear coordinates, of generalized Born-Oppenheimer type, and two Kohn-Sham 
Bogoliubov-de Gennes-like equations for the electronic states. Each of the equations 
involves its own exchange-correlation potential that requires approximation. 

In the standard approximation invoking Migdal's theorem the first equation describes 
fully dressed harmonic phonons. The formalism is extended to non-zero temperature 
thermal equilibrium, finally providing properties of the superconductor below T$_c$. Above the 
calculated T$_c$, the order parameter and corresponding potential vanish, and the 
solution reverts to the usual DFT description of the normal state. The schematic flow 
chart in Fig.~\ref{fig:scDFT1} provides an outline of the computational program.

Another central feature of SCDFT is that the empirical Coulomb parameter 
$\mu^*$ no longer appears. 
Instead, the screened Coulomb interaction is calculated from a functional involving the 
electron dielectric function $\epsilon(\vec q,\omega)$ as the central quantity. 
A few choices 
for this functional have been tested, analogous to the DFT exchange-correlation functional 
that likewise has
undergone refinements over several decades. 

\subsubsection{Tests for elemental metals}
This formulation with its computational implementation (see the flow chart in 
Fig.~\ref{fig:scDFT1}) has been shown to be remarkably accurate for elemental 
superconductors, especially considering the lack of any empirical input. In the 
upper panel of Fig.~\ref{fig:scDFT2} the calculated value  of T$_c$ is compared with 
the experimental value for five elements with very different strengths $\lambda$ and shapes 
of $\alpha^2F(\omega)$, all within the 1-10K range of T$_c$. For the 
``TF-FE'' choice of screening 
functional (for details see the original papers \cite{SCDFT1,SCDFT1b,SCDFT2}), there is 
negligible error (the maximum is 0.7 K for Ta). This accuracy for low to 
modest T$_c$ superconductors is impossible 
when $\mu^*$ is uncertain ({\it i.e.} is used as an empirical parameter) because the 
difference $\lambda-\mu^*$ involves both the  uncertainty in $\lambda$ (from 
approximate functionals and lack of full numerical convergence) and the fundamental 
uncertainty in $\mu^*$, while the value of T$_c$ is approaching zero because 
$\lambda$ is approaching the vicinity of $\mu^*$.  

The lower panel Fig.~\ref{fig:scDFT2}(b) illustrates a different test of 
the theory. It displays the calculated $T$=0 static gap $\Delta(0,0)\equiv \Delta_0$ 
compared to the experimental value. The correspondence is somewhat off for the transition 
metals Ta and Nb, depending on the approximate screening functional` that is used, but the 
theory can be improved with better screening functions, likely at a cost in computational 
effort. The report \cite{SCDFT2} of these results indicates how the shapes 
and magnitudes of the electron-phonon spectral function 
are very different for Pb than for Al, Nb, or Ta, yet the theory is impressively accurate for all. 

\subsubsection{Application to intermetallic compounds}
In the design of and search for new superconductors, improved numerical efficiency is 
much desired. With such alterations made to the functionals (sometimes with concepts 
borrowed from models), the procedure was applied by Sanna, Pellegrini, and Gross to 
elemental metals\cite{SCDFT3}~including those mentioned above, to the transition metal 
carbide TaC and nitride ZrN, to intercalated graphite CaC$_6$, 
to the A15 compound V$_3$Si, 
to the `high T$_c$' boride MgB$_2$, and to the compressed hydride SH$_3$. For these 
compounds the results are much improved, sometimes dramatically (except for ZrN,
with T$_c$ predicted 3 K too high) 
compared to the 2005 implementation applied to the same compounds. The variation
of T$_c$ in this set of materials ranges from 1 K for Al to 203 K for SH$_3$, 
and a range of calculated $\lambda$ from 0.4 to 2.8. A likely area of needed 
improvement was identified as the functional related to Coulomb repulsion --
what one might heuristically associate with $\mu^*$.

Due to the stated `low computational cost' of the changes allowing substantially 
improved numerical efficiency, additional capabilities have become available. The full 
$k$-dependence of the gap over the Fermi surface, a property of increasing interest 
especially after the discovery of the strong multigap character of MgB$_2$, 
is one such capability. 

The point  of this subsection is to emphasize that a fully {\it ab initio} and 
unusually accurate calculation of superconducting state properties and T$_c$ has 
become available for electron-phonon superconductors. 
The existence proofs of the underlying 
formalism allows for extended 
functionals in which the electron-electron interaction plays a more active role 
in the pairing mechanism.

\subsection{Limitations of SCDFT/DFT-Eliashberg theory}
SCDFT is formally exact for a condensed matter system in thermal equilibrium, analogous to the formal exactness of diagrammatic perturbation theory. In practice one relies on (1) three functionals, most obviously that of DFT for the static lattice, then those necessary for the superconducting state, and (2) the Migdal theorem, which states that electron and phonon propagators can be treated in the single bubble approximation -- vertex corrections (more involved diagrams) are negligible. SCDFT involves approximations of the various functionals that appear when treating pairing.
 
\subsubsection{Anharmonicity; non-linear EP coupling; quantum proton}
For metallic EP superconductors, the theory outlined above has reached a level 
comparable to that for metals in the normal state (though not yet widely applied), 
whose phenomenology was formalized in Landau Fermi liquid theory. If the metal is not a Fermi liquid, which can be due to low dimensionality, disorder, or strong electronic interactions, these effects must be built into the functionals. Complications include anharmonicity 
of the phonons and non-linear EP coupling, both of which arise diagrammatically
as vertex corrections but are not  often approached strictly diagrammatically.
An example is anharmonicity, which can be 
folded into a quasi-harmonic treatment. 
Quantum behavior of the nuclei is 
more prevalent for hydrides due to the light proton mass. Effects on 
T$_c$ are material dependent; while corrections to T$_c$ from each of 
these effect are at
the $\pm$5-8\% level, and further can be either additive or competing, their impact 
can be more far-reaching in materials with a complex unit cell.

\begin{figure}[tbp]
  \hskip -4mm
  \includegraphics[width=0.97\columnwidth]{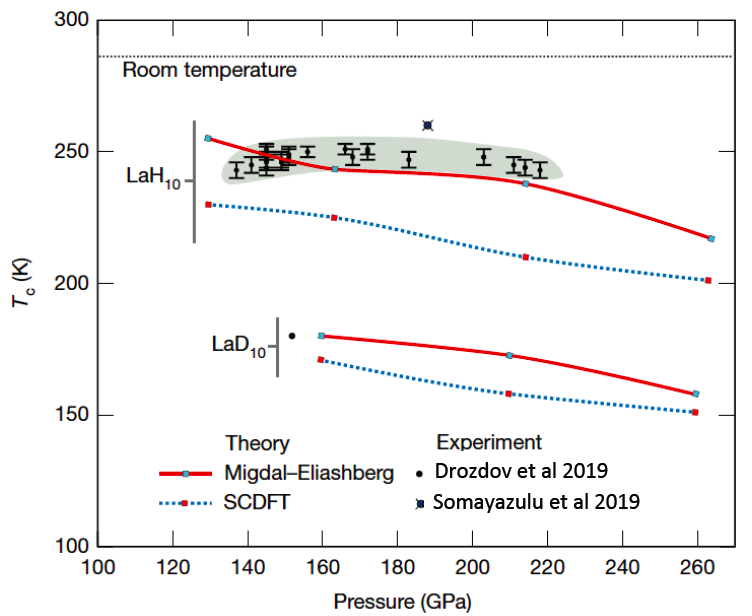} 
 \vskip -3mm
  \caption{Plot of calculated versus experimental values of T$_c$
 in $Fm{\bar 3}m$ LaH$_{10}$ and LaD$_{10}$. 
The measured values are shown as dots from \onlinecite{Drozdov2019},
and as dots with crosses from \onlinecite{Soma2019}.
The solid (red) lines are from an anisotropic Eliashberg calculation,
and the dashed {blue} lines are from SCDFT. Both include anharmonic
phonons within a stochastic self-consistent harmonic approximation. 
Adapted from \onlinecite{Errea2020} with permission.}
  \label{fig:scDFT3}
\end{figure}

One extreme example of the impact of quantum nature of the proton on
structure is provided by LaH$_{10}$ with its clathrate-like H cages 
around La.   
Using a classical description of the proton the $Fm{\bar 3}m$  structure becomes
dynamically unstable below 220 GPa, while it is observed to be a 
250-260 K superconductor around 200 GPa and somewhat lower (see Sec. VI
for more details and references). Errea {\it et al.} found that the
enthalpy surface contained several atomic configurations  \cite{Errea2020} 
with a lower enthalpy than the $Fm{\bar 3}m$ structure, inconsistent with
the observed structural data. Including the zero-point energy of quantum 
fluctuations of the proton along with anharmonic corrections,
the enthalpy surface reverted to a single minimum at the $Fm{\bar 3}m$
structure down to 130 GPa, restoring agreement with experiment. 

The fully corrected SCDFT results are shown in Fig.~\ref{fig:scDFT3}, along
with those from more conventional Eliashberg theory but with the quantum
proton and anharmonic corrections included so LaH$_{10}$ remains stable
down to 130 GPa, as found experimentally. The two theory curves differ
by about 10\% for the hydride and 5\% for the deuteride, where anharmonic
and quantum fluctuation effects would be smaller for the deuteride. 
For this case, it seems
that current SCDFT is not as accurate in reproducing the experimental 
results,\cite{Drozdov2019,Soma2019} which might indicate needed improvement
in the SCDFT functionals or might even be accidental given possible
complexities of sample preparation. The comparison in Fig.~8 is with
experimental results from two groups.\cite{Drozdov2019,Soma2019} 

\subsubsection{Eliashberg theory at strong coupling}
Over many years there have been studies following the expectation that strong EP coupling 
alone provides a breakdown of Eliashberg theory for the normal state. 
The question needs to be specified more
specifically, since the DFT-based Eliashberg theory is a given formalism. In the absence of
divergences, it makes specific predictions for the system, and these are what current and
past solutions describe.

The implicit question is more likely this: at strong coupling, does 
Migdal-Eliashberg theory give 
correct results for known or realistically predicted materials?  
Conclusions 
   (see for example \cite{gunnarsson}, \cite{chubukov}, \cite{oppeneer}) for 
treatments and references)
have often been based on the Holstein model, originally put forward 
as a minimal model treatment of EP coupling,
applicable to very low density carrier systems, {\it i.e.} semimetals and doped 
semiconductors. 

In the Holstein model, the EP interaction is entirely on-site: 
electron charge on a site is coupled to a non-specific scalar displacement 
on that site. Intersite 
coupling is due only to indirect coupling through the conserved charge
density; there is no modulation of an intersite hopping parameter. 
As mentioned in Sec.~II.E.3, Allen and Dynes provided study of T$_c$ 
in Eliashberg theory to very large
values of coupling, finding a smooth T$_c$($\lambda$) relation with no
evidence of a phase transition or even a crossover.
The question is: at very large coupling, does Migdal-based Eliashberg theory
cease to provide the behavior of real systems, and if so, how does this
happen?

DFT-based Eliashberg theory treats the 
force on an atom due to its displacement as a collective effect, 
involving changes in potential that induce forces from neighboring shells that 
impact, sometimes strongly, dispersion
of the renormalized (physical) phonons. An electron mass enhancement fixed 
to the Fermi energy results in an enhancement of the fermion excitation
density of states by a factor 1+$\lambda$. The resulting phonon
dispersion determines the
possible instabilities of the lattice.  Numerically exact (Monte Carlo) 
and other non-perturbative techniques cannot handle 
the resulting complications of real materials in currently available computational time. 
Put another way, the strong EP coupling limit based on the full crystal
Hamiltonian of Eq.~\ref{eqn:hamiltonian} has yet to be attempted. 

Experience up to the present, both experimental and computational, indicates that 
the primary limitation of DFT-Eliashberg theory derives not from 
strong coupling {\it per se}, {\it i.e.} electron mass enhancement,
but rather from lattice instability sometimes already occurring at moderate coupling. 
There are numerous examples indicating that the strong coupling 
related lattice instability, 
which is strongly material-dependent and often Fermi surface related, and can occur 
already at moderate coupling. Phonon branches are renormalized to lower energies 
in a momentum- and branch-dependent fashion, followed by zero and (calculated) 
imaginary frequencies as coupling strength increases. Calculations up to 
$\lambda$=4 (see the following sections) indicate that it is possible 
to have standard EP coupled 
systems at that coupling strength (though lattice instability is imminent).

\subsubsection{Inapplicability of Migdal's Theorem}
We return briefly to the more formal limitations of Migdal-based Eliashberg theory. 
At the most basic level, the regular decrease in importance of successive
terms in the perturbation expansion relies (in a 
self-consistent treatment at a chosen number of terms) on the 
smallness of a ratio of phonon velocity to Fermi velocity
\begin{eqnarray}
\frac{\omega_q/q}{v^*_k} =\frac{\omega_q/q}{v_k/(1+\lambda_k)} << 1,.
\end{eqnarray} 
often stated heuristically in terms of the electron-to-ion mass ratio $\sqrt{m/M}$, around 1/300 
for medium mass ions.
The ratio of phonon phase or group velocity to Fermi velocity is typically 
10$^{-2}$ or smaller in Fermi liquid metals, providing an excellent expansion parameter. 
The renormalized (downward, by the EP mass enhancement $1+\lambda_k$) Fermi surface 
velocity $v^*_k$ allows for a self-consistent treatment at second order perturbation theory. 

In a given band this velocity might approach zero near E$_F$ due to a van Hove singularity, 
but typically in a small region of the zone and for a single band out of several, perhaps
giving rise to unusual effects but not an invalidation of the theory. The platform of 
flatband materials provides possible realizations for violations of this condition, 
but the small phase space in a conventional metal, viz. a wideband compressed hydride
(compressed hydrides have occupied conduction bandwidths or 25 eV or more), 
argues that such occurrences will be rare and, in a many-band background, of minor impact.  

Fermi surface nesting, which has an extensive literature, can exaggerate a related 
condition of this sort, providing a 
larger phase space in which inter-Fermi-surface scattering acquires a significant 
phase space. This possibility is real and the resulting behavior of the system (several 
possible instabilities involving broken symmetries) is material- (and model-)dependent.   

\subsubsection{Thermal fluctuations at high T$_c$}
Superconductivity theory, especially in the strong coupling regime, posits a complex-valued
order parameter (related to the energy gap function). 
In bulk EP superconductors fluctuation of the
magnitude of the gap is rarely regarded as a limiting factor for high T$_c$. In the 
quasi-2D high T$_c$ cuprates, fluctuations of the phase of the order parameter has
received a great deal of attention, with one viewpoint being that fluctuations are
especially large, and possibly limiting, in the pseudogap region of the 
doping-temperature phase diagram. While fluctuations are much less of a factor in
three dimensions than in two, the elevated values of T$_c$ in compressed hydrides
invites consideration of the question. 

An energy scale for the phase stiffness was given by Emery and Kivelson as
\bea
V_o = \frac{(\hbar c)^2 a}{16\pi e^2 \Lambda_L^2(0)}
\eea  
in terms of the London penetration depth $\Lambda_L(T)$ at zero temperature and 
with characteristic
length scale $a$ which would be comparable to the superconducting coherence length.
If $V_o$ (in temperature units) begins to approach T$_c$, fluctuations would begin
to arise, requiring a generalization of Eliashberg theory where no fluctuations 
appear (it is a mean field theory in this respect).

Eremets and collaborators \cite{EremetsData,Minkov2022a,Minkov2022b} 
have provided, from upper and
lower critical field measurements, the necessary data 
for SH$_3$ and LaH$_{10}$, with T$_c$
around 200 K and 260 K respectively. The values are, respectively, $a\approx \xi$
= 1.84 (1.51) nm, $\Lambda_L(0)$ = 18.2 (14.4) nm. A more recent design for 
measurement of flux pinning, for different samples of SH$_{3}$, gave
$\Lambda_L(0)$=27 nm. In each compound the ratio
$V_o/T_c$ on the order of $10^3-10^4$, so order parameter phase fluctuations are  
negligible in compressed hydride superconductors. 

\subsubsection{Broader comments}
The conclusion based on the full DFT-Eliashberg theory and behavior 
of real materials is 
that there is currently no established limit on EP-mediated 
superconducting T$_c$, if
the aforementioned vertex corrections are taken into account. A few 
calculations are predicting above room temperature superconductivity 
in certain compressed hydrides where the strongly coupled 
phonons have very high frequency and $\lambda$ exceeds four (see below). While there 
is no known fundamental limit, it does not follow that prospects 
are rosy. However, the high 
T$_c$ regime (versus the more specific but elusive large $\lambda$ regime) remains 
open to new discoveries. 

The challenge is (i) to retain lattice stability and (ii)
achieve strong coupling to high frequency phonons: both high $\Omega$ (an appropriate
phonon frequency scale) and large $\lambda$. Very high T$_c$ seems to require a
T$_c$ expression that is increasing in both $\Omega$ and $\lambda$.  The theory
gives the  Allen-Dynes 
conclusion: the regime of very high $\lambda$ is 
\begin{eqnarray}
T_c \propto \sqrt{\lambda \omega_2^2}=\sqrt{\eta/M}. 
\end{eqnarray}
Somewhat peculiarly, $\eta=N_{\uparrow}(0){\cal I}^2$ is the purely electronic 
quantity shown by McMillan
to be the vibration frequency
independent moment of $\alpha^2F$. This result 
is subject to vibrational behavior by only requiring no unstable modes, but an
otherwise arbitrary vibrational spectrum. The inverse square root of 
mass factor is in line with
Ashcroft's original proposition \cite{Ashcroft1} -- 
hydrogen systems might have an order of magnitude times larger
limit than Nb compounds (with other differences depending on $\eta$).  Worth repeating:
the regime of very high T$_c$ may not be the same as that of very large $\lambda$. 
Hydrides suggest that very high T$_c$ is the regime of very high frequencies.
Unhappily, it is difficult to imagine higher frequencies except at even higher pressures. 

\section{Applying the theory}

With DFT-Eliashberg theory in hand, the periodic table and compilations of known 
superconductors provide an imposing number of possible applications
posing the question: 
which materials systems are the most rewarding for study. This question is discussed 
briefly in this section, with the question of how to sample the many possibilities 
giving some discussion in the following section. 

\subsection{Choice of Favorable Materials Platforms}
Regarding ever higher T$_c$, transition metal based materials had been sampled and 
studied, experimentally and theoretically, by 2010, and promising directions 
were few, but included organic superconductors and interfacial (excitonic) 
superconductivity.. 
Considering possibilities more generally, the favored palette of atomic constituents 
for higher, possibly room temperature, superconductors had already been presented  
by Ashcroft. In 1968, while research focused on transition metal 
compounds, he proposed elemental metallic hydrogen as a very high T$_c$ 
material, \cite{Ashcroft1} based on (1) its small mass favoring high frequencies, and 
(2) the vibrating proton without core electrons, which if not too strongly screened 
should provide the strong scattering that would be required for pairing at high 
frequencies. 

These properties were supported by the BCS Eq.~\ref{eqn:BCS}, although 
little of a material-specific nature was understood relating to higher T$_c$. 
Ashcroft understood that it would require high pressure to metalize hydrogen, 
specifically to break the H$_2$ molecular bond to create an ``atomic hydrogen metal'' 
and bring strong scattering processes to the Fermi energy. That it would require 
of the order of 1 TPa (ten megabar) or more \cite{Ceperley} 
to break the H$_2$ bond may not have been anticipated.

The theoretical and computational capabilities to address this question would not be available until the next century. Skipping ahead: in 2004 Ashcroft refocused his concept, arguing
\cite{Ashcroft2} that hydrogen-rich molecules, viz. H$_2$, CH$_4$, NH$_3$, etc., could circumvent the challenge of breaking the strong H$_2$ bond by replacing it with a weaker bond, and also provide ``precompression'' (higher H density) in the experiment, thereby lowering the required pressure to produce what would be essentially metallic hydrogen.  Section V reveals how remarkably successful this path has been.

\subsection{Computational Implementation}
The formalism and some important analysis was complete by the mid-1980s, where self-consistent DFT calculations could provide the electron wavefunctions and band structures giving Fermi surfaces in excellent agreement with experiment, and phonons were becoming available but were still challenging computationally. The expression for the Eliashberg function, re-expressed from Eq. 3, illustrates one of the computational bottlenecks:
\bea
\alpha^2F(\omega)=N_{\uparrow}(0) 
 \frac {\sum_{k,Q}|M_{kk'}|^2 \delta(\omega-\omega_{Q})    
   \delta(\varepsilon_k)\delta(\varepsilon_{k+Q}) }
  { \sum_{k,Q}\delta(\varepsilon_k)\delta(\varepsilon_{k+Q})} \nonumber
  \label{eqn:a2f2}
\eea
where the Fermi energy $\varepsilon_F$=$0$, and $Q=k'-k$ is the wavevector of the 
phonon scattering an electron from state $k$ to $k+Q$, both on the Fermi surface. 
Necessary sums over bands 
and phonon branches are implicit. 

The sums that are shown are each over the three 
dimensional Brillouin zone, confined by the pair of $\delta$-functions to lines 
of intersection of Fermi surfaces, one displaced by $Q$ from the other and requiring 
fine meshes for convergence. As before, the density of states factor is for a 
single spin, designated by the arrow on $N_{\uparrow}(0)$. This expression makes 
evident the geometrical interpretation of a double average of $|M^2|$ over the 
line of intersection of two Fermi surfaces with relative displacement $Q=k'-k$, 
all done in a frequency $\omega$-resolved fashion.

To emphasize the numerical challenge, we note that the EP matrix element 
$M_{k,k'}$ of Eq.~\ref{eqn:matrixelement} 
requires the computation of the self-consistently screened potential due to each 
phonon displacement, then requiring identification and evaluation of the 
matrix element between 
electron states $k$ and $k+Q$ on the Fermi surface, with necessary band and 
phonon branch indices.  These are included within the six dimensional integral.
This extensive computation has, through innovative 
algorithms, been brought to viable although still time-consuming level. The 
electron states are expressed in terms of localized Wannier 
functions, \cite{Wannier} and the phonons are expressed in terms of localized 
lattice Wannier functions. \cite{latticewannier,cockayne} This combination 
considerably speeds the various required zone samplings. 

Linear response 
algorithms have been implemented to enable the phonon-induced change in 
screened potential to be calculated from the formal 
infinitesimal-displacement limit.  
Beyond the basic DFT formalism and codes, evaluating $\alpha^2F$ to convergence 
required a sequence of advancements of formalism and construction of codes, many of 
which were adapted to parallel computation. In lieu of attempting to describe them, 
we refer the reader to a modern treatment of anisotropy by Margine and 
Giustino, \cite{Margine}, an extensive review article on the EP formalism with 
several of the algorithms and references, \cite{Giustino} and a 
monograph on materials modeling \cite{GiustinoBook}.

\section{Crystal Structure Prediction}


\subsection{General recent activities}
The previous section described the developments that have led to the current capability: given a dynamically stable specific compound, the EP coupling strength $\lambda$, superconducting T$_c$, and several properties of the superconducting state can be calculated reliably. Design of new superconductors requires a separate capability: the prediction of new stable crystalline materials. 
Design and discovery of new materials was an occasional occurrence until the several agency-wide U.S. program {\it Designing Materials to Revolutionize and Engineer our Future} (DMREF) that was initiated in 2011. Related programs have emerged in other countries.  

The idea  was to push the ever expanding computational power, and theoretical and 
algorithmic development, to design new materials and properties in many classes, 
to accelerate experimental discovery, then to speed time to market of new products. 
Many new programs have supported this initiative, which emphasized computational 
theory-experiment partnership and research, development, and industry synergy.

Materials design, even restricting oneself to crystalline materials with modest sized 
unit cells, is a challenging process. First is the choice of number of elements 
(we discuss binaries; ternaries and beyond require thoughtful choices) and their 
stoichiometry, with some chosen property in mind but initially in the background. 
A great deal is known about structural phases of elemental materials, and about 
the concentration-temperature phase diagrams of  several binary compounds, and 
certain ternary classes at ambient pressure. Given the periodic table of 100 elements 
($X$,$Z$), there are $\sim 5000$ X-Z pairs, dozens of stoichiometries, and for each, 
as many as dozens of reasonable structures. For a brute force search, a DFT relaxation 
would require perhaps 2.5$\times 10^5$ compounds, 
certainly a daunting task. Classification into the 230 space 
groups can help to organize the exploration, but samples with first order transformations 
readily skip over space group symmetries. 

The focus in the following is on Ashcroft's suggestion of H-rich materials, which for binaries reduces the search drastically to ${\cal M}$H$_n$ compounds, but leaves 
a still challenging task (here the formula is normalized to one ${\cal M}$ 
atom). The focus here is compressed hydrides, 
adding the essential pressure variable to the space to be explored. The challenge is to identify candidates that are metallic and thermodynamically stable (metastable phases that are not far from stability may be of interest), then check whether they are dynamically stable. Finally, calculation of electron bands and wavefunctions, and the phonon spectrum, followed by $\alpha^2F$. Then calculation of T$_c$, $\Delta(\omega,T)$, and a few other properties of the superconducting state can be carried out, using the algorithms and codes mentioned above.

\subsubsection{Free energy functional}
To identify thermodynamic stability (compounds that will not decompose into two or more phases with lower energy) for any chosen pressure P and temperature T, an efficient numerical scheme is required due to the computational complexity of the exercise. Fundamentally, the goal is to identify, for a given stoichiometry ${\cal M}$H$_n$, the minimum enthalpy H(P,T) over the possible crystal structures. Somewhat more precisely, one really needs for the most precise prediction the minimum of the Gibbs free energy 
\begin{eqnarray}
{\cal F}(P,T)&=& \left [E(P,T) +PV(P,T)\right ] -TS(P,T) \\ \nonumber
             &\equiv& H(P,T) -TS(P,T),
\end{eqnarray}
in terms of the internal energy $E$, enthalpy $H$, volume $V$, 
and entropy $S$ (mostly lattice vibrational at temperatures of interest). 

It is found that the lattice zero-point energy in $E(0,0)$ is more important for hydrogen than for heavier atoms, and the entropy term can shift phase boundaries. The zero point energy and entropy requires calculation of the phonon spectrum, but much screening of candidates can be done without this step. The volume $V$ and internal structural parameters are relaxed (at $T$=0) to get the enthalpy $H(P,0)$, and the vibrational entropy is calculated when that level of precision is desired. When $H(P,0)$ is plotted versus concentration of H 
the resulting curve is called the convex hull. The minimum gives the predicted most stable stoichiometry, which is calculated for each pressure of interest. 

In the area of compressed hydrides, it has become fairly standard to calculate the complex hull for each pressure of interest, and very often to check the most favorably case(s) for dynamic stability. That information makes it worthwhile for the experimenter to attempt synthesis and characterization, and this symbiosis is evident in several productive collaborations.

\subsection{Evolutionary prediction}
An enabling capability in the structural search has been the development, in this century, of evolutionary crystal structure prediction. There are a few methods in use, but the concept is to choose a few candidate structures, relax them, and compare properties, especially the total energy, or under pressure, the formation enthalpy. The most favorable candidates are chosen to guide the construction of new candidates -- typically, derived ``evolutionarily'' by some algorithm -- until the most favorable candidate is obtained (for a given pair and a given cell or supercell size, as a practical limitation). 

Crystal structure prediction progressed from early random sampling, basin-hopping, and force-field molecular dynamics to first principles DFT-based enthalpy comparisons, often outlining the convex hull of thermodynamic crystal stability. Modern methods, with some prominent ones (with clues to methods contained in their acronyms) being CALYPSO \cite{CALYPSO} (Crystal structure AnaLYsis by Particle Swarm Optimization),  USPEX \cite{USPEX} (Universal Structure Predictor: Evolutionary Xtallography), AIRSS \cite{AIRSS} (Ab Initio Random Structure Searching), and XtalOpt \cite{XtalOpt} (crystal structure prediction  and optimization). These codes  incorporate various algorithms from simulated annealing, evolutionary/genetic algorithms, minima or basin hopping, particle swarm optimization, metadynamics, and (quasi)random searches, \cite{Zurek2022b} to search the necessarily broad configuration space. A monograph can be consulted for further information \cite{oganov}.  

\begin{figure*}[tbp]
  \hskip -4mm
  \includegraphics[width=1.905\columnwidth]{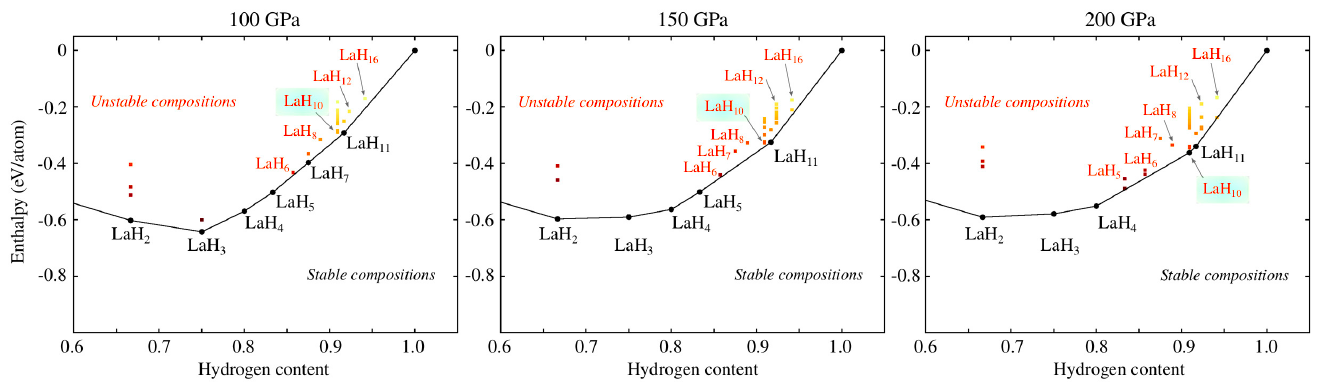}
  \includegraphics[width=1.755\columnwidth]{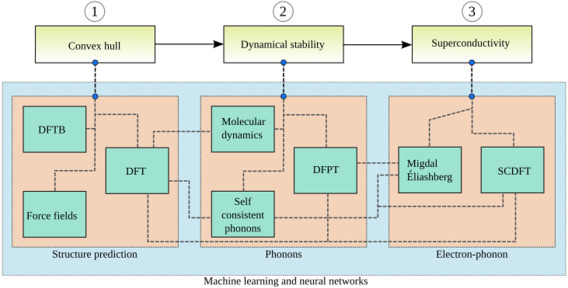}
  \caption{Top row of panels.  
   For the LaH$_n$ system, the computed convex hull, with 
  specific calculated formation enthalpies with respect to elemental La and H$_2$. 
  Shown in the three top panels are pressures of 100, 150, and 200 GPa for  
  the stoichiometries that were obtained in the search. 
  LaH$_{10}$ is the compound of most interest, marked with light green 
   background (or gray in grayscale print). 
  Convex hulls often result in several compounds near the 
   convex hull, and may be obtained from hundreds or thousands of 
   enthalpy calculations. Adapted from Errea {\it et al.} with permission.
     Bottom row of panels.
     A schematic diagram of the progression of the high T$_c$ search program. 
   First, the thermodynamic stability is obtained from compounds on the convex hull.
   Second, dynamic stability is studied, and phonon spectra are obtained. 
   Third, the DFT-Eliashberg calculation: EP coupling and the phonon spectrum 
    to obtain $\alpha^2F(\omega)$, and T$_c$ is obtained from the 
    Eliashberg equation or from one of the fit expressions 
     T$_c$($\lambda,\mu^*,\omega_{log},\omega_2$). DFTB, DFPT, and SCDFT 
     indicate separate DFT-based capabilities. Adapted from 
     Flores-Livas {\it et al}  with permission.
  }
  \label{fig:4panels}
\end{figure*}

Actual procedures differ, and the likelihood of finding the true stable formation enthalpy minima is constrained by a few factors, such as unit cell size (faced by all methods) and the effort spent to sample the full phase space, which as mentioned is a daunting task. 
A typical procedure might proceed as follows. 
After choices of metal atom, H concentration $n$, and some candidate structures (or space groups), the volume and atomic positions are relaxed subject to the chosen pressure. In early steps faster methods with somewhat lower precision can be used.  Evolutionary steps based on the most favorable candidates produce new candidate structures. 
The candidate structures are combined, using various algorithms, to produce the next
level candidates.

The process is continued (a computationally taxing procedure) until negligible 
improvement is occurring, signaling convergence.
The H concentration $n$ is changed and the process repeated; sometimes 
this is automated. 
Varying $n$, which for a given system can involve hundreds of thousands of 
structures, \cite{Zurek2022b} the ``convex hull'' delineating the stable compounds 
in this binary system can be plotted. Examples of the convex hull of LaH$_{10}$ at 
three pressures are shown in Fig.~\ref{fig:4panels}(a). 
This method, involving enthalpy calculated from DFT, has proven successful in 
predicting stable candidates. 
Various chosen pressures must be computed independently.

Thermodynamically stable candidate structures then must be checked for 
dynamic stability (absence of imaginary frequencies) to be valid 
predictions. Stability can be determined either by phonon spectrum 
calculations or by {\it ab initio} molecular dynamics. Only then can 
DFT-Eliashberg theory be applied to obtain T$_c$ and a selection of 
other desired properties obtained from the gap equation. The procedure 
for full search for high T$_c$ is outlined in schematic form in 
Fig.~\ref{fig:4panels}(b).

\subsection{Machine learning, data mining}
As in many of the sciences and elsewhere, machine learning techniques are being 
applied to materials design, but few yet with applications to compressed hydrides. 
The basic idea is pattern recognition: provide a large database to give the neural 
net the opportunity to identify certain characteristics (``training'') that are 
(statistically) related to given descriptors. Application of the trained
apparatus to new possibilities produces likely candidates 
for the desired characteristics (viz. T$_c(P)$) with rectitude estimated by various 
statistical measures. While earlier applied to address other materials properties, 
applications to general superconducting materials have been implemented 
only more recently. 
A large database of known superconducting materials exists, consisting of the compound 
formula, crystal structure, and T$_c$.  However, no experiment-derived database for 
compressed hydrides exists due to the dearth of hydride examples, so training must be done 
on computationally predicted cases. 

In 2017 an example of machine learning related to superconducting T$_c$ was provided by
Stanev {\it et al.} \cite{Stanev2018} Their study covered a large range of values of T$_c$
and the SuperCon database of over 12,000 superconductors 
to provide training. 
\footnote{\label{SuperCon}DICE: a data platform for materials science.
National Institute of Materials Science, Materials Information Station,
SuperCon, http://supercon.nims.go.jp/index\_en.html (2011).}
With additional
guidance of materials properties from the AFLOW repository, \cite{AFLOW} 
the procedure was applied
to the entire Inorganic Crystal Structure Database. \cite{ICSD} More than 30 non-copper and 
non-iron materials, with all being multicomponent oxides, were identified as the most 
promising candidates.

The machine learning study of hydrides by Hutcheon {\it et al.} provides a more
specific instructive example. \cite{Hutcheon2020} For descriptors for candidates $M$H$_n$, 
they chose H content ($n$), and the metal element ($M$) size (van der Waals radius), 
atomic number, mass, and electronic configuration (number of $s, p, d, f$ electrons). 
Notice that these descriptors have little direct relation to the quantities that 
determine $\lambda$ or phonon frequencies or interaction strength. Their result identified 
the first three columns of the periodic table as best candidates for $M$, a feature 
also noted in various hydride overviews. 

These low electron affinity elements give up much or even all of their valence electrons to the H sublattice. This added charge raises the H $1s$ occupation above the half-filled level, promoting metallicity.  
In studies of atoms $M$ across the entire periodic table, a few outliers exist: atoms with open $d$ or even $f$ shells. Interestingly, the first discovery, SH$_3$, is a different sort of outlier, with sulfur's open, roughly half-filled $p$ shell. 
Also unusual for SH$_3$ is the sharp van Hove singularity at the Fermi level,
which could contribute to it being an outlier. \cite{Quan2019}

\section{The breakthrough discoveries:
  theory then experiment}
  
\subsection{High pressure experimentation}
According to Ashcroft's original concept, \cite{Ashcroft1} metallic atomic hydrogen should provide the acme of T$_c$, since the atom has the smallest mass, with highest frequencies if force constants are strong, and the potential for strong scattering of electrons by proton displacements that seemed likely to Ashcroft. 
Calculations sometimes including the quantum nature of the proton have predicted the stable structures versus pressure.  The predictions are T$_c$ of 500K or higher, \cite{Ceperley} requiring pressures of 500 GPa or higher. 
Few attempts have reached a static pressure this high, so Ashcroft's second suggestion \cite{Ashcroft2} has been the avenue of choice: use H-rich molecules as the ambient pressure sample to avoid strong H$_2$ bonding and antibonding states being pulled away from the Fermi level, and to exploit precompression (higher H concentration and density).  

Another point merits mention. Over past decades there have been reports of signals of possible room temperature superconductivity, usually in resistance or susceptibility measurements, which are the most straightforward evidences of superconductivity. 
The samples were invariably polycrystalline, multiphase, or disordered to the point of amorphous. 
Transport and magnetic measurements often show anomalies in such samples. When such signals are not reproducible, they have made the community skeptical to the point that ``USO'' is a recognizable acronym -- unidentified superconducting object. It is possible that some of them could be evidence of interface superconductivity or some other unusual type, but if not reproducible a report does not receive extended notice. 

For this reason the discoveries below focus on reproducible results, noting 
confirmations. It must however be recognized that the samples in diamond anvil cells 
are far from the ideal single crystals that are often available at ambient pressure. 
The compounds are synthesized within the tiny pressure
cell at megabar pressures, and temperatures up 
to 2000K are varied. Resulting samples will typically be polycrystalline, strained, and 
possibly multiphase, and with hard-to-determine stoichiometry. 
For first order phase transitions the free energy barriers may be high, making it 
challenging to reach certain Gibbs free energy minima. 
For these reasons various confirmations of these first three discoveries of 
approaching or near room temperature superconductivity will be noted. Also, given 
the complexities of samples (mentioned above) at a given P and T, experimental data 
will not be uniform across groups, nor even across a given laboratory's runs. Thus ``reproducibility'' and comparison with theoretical predictions should be interpreted accordingly.

\subsection{SH$_3$: the initial breakthrough} 
\subsubsection{ Theory} 
In 2014 Li {\it et al.} \cite{LiMa2014} employed the CALYPSO structure prediction code \cite{CALYPSO} to identify candidate structures with SH$_2$ stoichiometry at high pressure. Note that this formula is isovalent with H$_2$O, and the molecules are isostructural. The much more strongly bonded H$_2$O is known to remain insulating up to the current highest static pressure available. Their study, finding insulating structures at lower pressures, focused on a transition between two predicted metallic structures in the 130-160 GPa region. The lower pressure compound has a low symmetry $P\bar{1}$ space group, polymeric structure, with an atomic H $Cmca$ phase with a layering of H and of S atoms occurring at higher pressure. The calculated T$_c$ was maximum at the transition ($\sim$160 GPa), around 60K in the $P\bar{1}$ structure and 82K for $Cmca$, and decreasing with pressure in this latter phase.   The maximum coupling strength, occurring at the structural transition, was $\lambda$=1.25, with 40\% attributed to S modes (which comprise 1/3 of the phonon branches).

\begin{figure*}[tbp]
  \includegraphics[width=2.2\columnwidth]{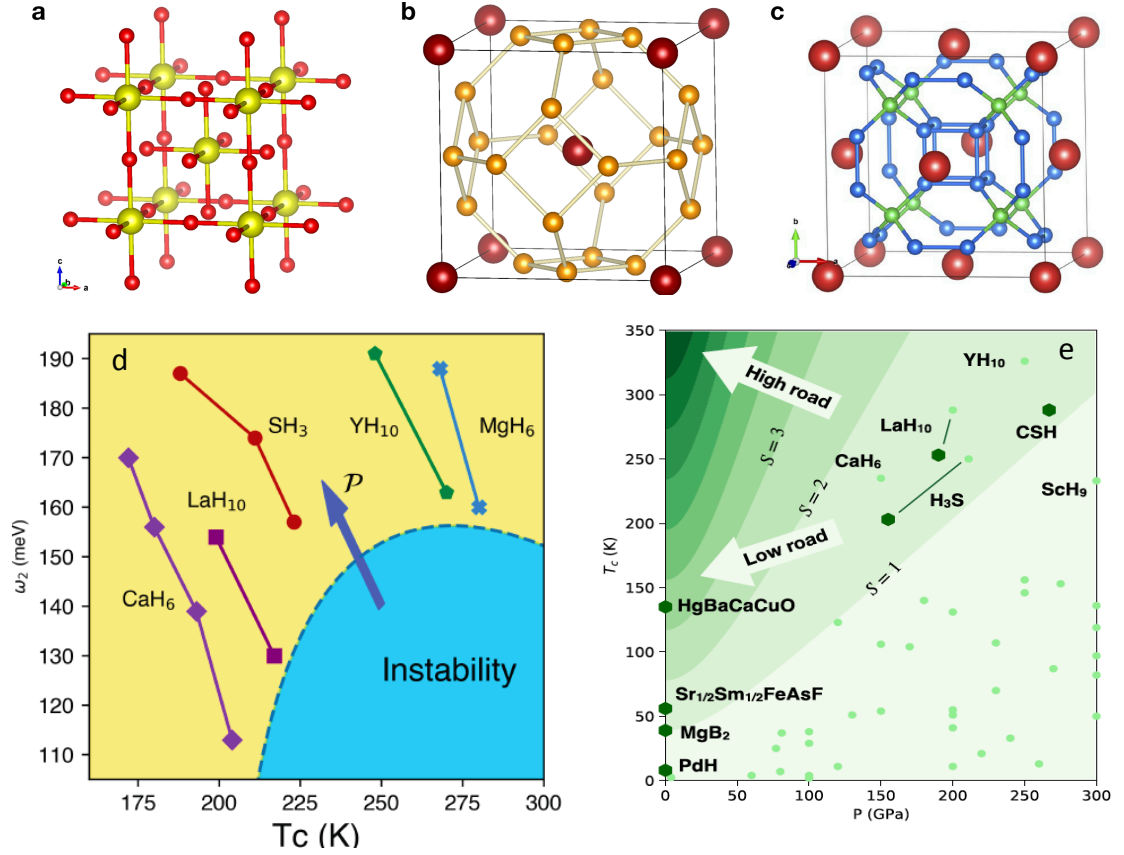} 
  \caption{(a-c). The structures of SH$_3$, YH$_6$, and LaH$_{10}$, respectively, 
  illustrating the high symmetry and the progressions toward sodalite or 
  clathrate-type structures with increased H content. Small circles indicate H
 atoms, larger circles denote metal atoms.
  (d) A schematic $\omega_2$--$T_c$  phase diagram of calculated results 
   for five compressed hydrides in three crystal structure classes shown above. 
  Note that pressure is increasing in the upper left direction, the blue region
  (lower right  region) denotes the low pressure region of instability of 
  the structures that are considered. 
   Solid lines connect values of a compound changing with pressure.
  Pressure lowers T$_c$; as pressure increases, $\omega_2$ increases but 
$\lambda$ and T$_c$ decrease. Lowering pressure increases T$_c$ (and $\lambda$)
but leads to dynamic instability. The trends suggest an ``island of instability''
for compressed hydrides. 
  (e) Pickard's scatterplot in the P-T$_c$ plane of experimental (hexagons) and theoretical (circles) positions for several superconductors. 
  The contours indicate values of the {\it ad hoc} figure of merit 
  $S=T_c(K)/[T_{c,MgB_2}(K)^2 +P(GPa)^2]^{1/2}$, 
  which focuses attention on the desirability of low pressure P for higher T$_c$ 
  (see Boeri {\it et al.}, Sec. 12.)}
  \label{fig:structures}
\end{figure*}

Appearing immediately after, also in 2014, was a prediction by 
Duan {\it at al.} \cite{Duan2014} that for a SH$_3$ stoichiometry 
[labeled at the time after the initial constituents, (H$_2$S)$_2$H$_2$), 
and later as H$_3$S], a high symmetry, atomic H structure pictured
in Fig.~\ref{fig:structures}(a) was predicted 
to be the most stable one in the 150-200 GPa range, and the calculated 
T$_c$ peaked just above 200 K near the lower pressure range of the 
160-200 GPa region. Coupling strength $\lambda$=2.2 was calculated, with 
25\% arising from the low energy S modes. The cubic structure has one of the highest symmetries possible for this stoichiometry: SH$_6$ octahedra in a body-centered
arrangement are connected on two sublattices by H atoms which are coordinated 
only with two S atoms, see Fig.~\ref{fig:structures}(a). Such a remarkably 
high prediction must have seemed to most readers as beyond comprehension, 
yet it was based on successful structure search algorithms and the established 
DFT-Eliashberg theory. Theoretical study of sulfur hydrides was confirmed and 
extended in 2015 by Akashi {\it et al.} \cite{Akashi2015} and 
Errea {\it et al.}, \cite{Errea2015} which promoted the understanding of the 
relation of electronic structure and microscopic processes to promote high T$_c$. 

\vskip 2mm \noindent 
\subsubsection{Experiment} 
Independently, experimentalists were working on S-H samples. Even in 
2014, Drozdov and collaborators \cite{Drozdov2014} posted a notice of 
pressure-induced T$_c$ up to 190 K in sulfur hydride samples, and in 
2015 the published announcement \cite{Drozdov2015} was T$_c$ as high as 
203 K around 160 GPa in the H-S system. The superconductor responded 
to magnetic field in the anticipated way, revealing the expected critical 
field H$_c(T)$ behavior versus temperature.  

The sample was later determined by Einaga {\it et al.} \cite{Einaga2016} 
to have the bcc S sublattice predicted by Duan {\it et al.}, and it was 
concluded, based also on computational input and experimental volume, 
to be bcc SH$_3$ as shown 
in Fig.~\ref{fig:structures}(a). The hydrogen$\rightarrow$ deuterium 
isotope shift of T$_c$ was large, as theory predicted, and T$_c$ versus 
pressure was reproduced. This discovery opened the field to the realistic possibility 
of room temperature superconductivity. High T$_c$ in this T-P regime has 
been confirmed, for example by Huang {\it et al.} who reported from s
susceptibility measurements T$_c$ as high as 183K around 150 GPa. \cite{Huang2019}
 
Magnetic measurements by Eremets and 
  collaborators \cite{EremetsData,Minkov2022a,Minkov2022b}
led to the superconducting material parameters $\Lambda_L\sim 18-27$ nm,
$\xi$=18.4. The resulting Ginzburg-Landau parameter 
$\kappa_{GL}$=$\Lambda_L(0)/\xi(0)$=10-15
indicates strong Type II superconductivity.

In retrospect not surprising, there was significant skepticism about SC at
200K until the following two higher T$_c$ discoveries were made and reproduced.
However, for those (relatively few) who understood the degree and accuracy
of the computational theory, the fact that {\it experiment agreed with theory} 
would have been convincing in itself, as it was for this author.

\subsection{LaH$_{10}$: approaching room temperature}
\subsubsection{ Theory }
The remarkable success of design and discovery for SH$_3$ emboldened 
the superconducting materials design community. Binary hydrides ${\cal M}$H$_n$ 
were the focus. Varying the metal ${\cal M}$ atom -- valence, size, chemistry -- 
and moving toward superhydrides ($n>6$, say), would seem to approach the 
optimal combination to move toward the idealistic case of Ashcroft's metallic hydrogen.  

In 2017 two theoretical groups, Liu {\it et al.}, \cite{HLiu2017,HLiu2018} and Peng {\it et al.}, \cite{Peng2017} nearly simultaneously predicted ${\cal M}$H$_{10}$, with isovalent ${\cal M}$=La and Y, to have high T$_c$, in the 275-325K range depending on element and pressure (always above 200 GPa). Liu {\it et al.} calculated $\lambda=2.2$, with T$_c$ around 265K. The structure again was the highest symmetry possible for this stoichiometry, cubic with a clathrate-like shell of 32 H atoms surrounding the metal atom on its fcc sublattice, pictured in Fig.~\ref{fig:structures}(c). Values for $\lambda$ were in the 2.2-2.6 range, similar to SH$_3$ but also similar to Pb-Bi-Tl alloys from the 1970s with maximum T$_c$=9K that is 35 times lower. \cite{AllDyn} The differences in T$_c$ are due to the very high H vibrational frequencies (Ashcroft's primary point) while retaining strong coupling to Fermi surface electrons.   

Further studies of this superconductivity were provided by 
Liu {\it et al.} \cite{HLiu2019} and the quantum (zero point motion) nature 
of the structure by Errea {\it et al.} \cite{Errea2020} as discussed in
Sec. II.G.1.
Theoretical work by Ge {\it et al.} \cite{Ge2021} indicated that doping 
LaH$_{10}$ by B or N on either the La or H site at the few percent level 
might raise T$_c$ by 30K, {\it i.e.} to T$_c\approx$290K, in the 
240 GPa regime. This doping also strongly tends to drive the alloys
toward lattice instability, a common occurrence when coupling is increased.

\vskip 1mm \noindent 
\subsubsection{ Experiment} 
The superconducting materials discovery (experimental) community was also stimulated by the developments on SH$_3$. 
Synthesis and evidence of superconductivity in lanthanum superhydride around 260K was announced in two publications \cite{geballe2018,Soma2019} in 2018-2019. Resistivity drops occurred in the 180-200 GPa range for various runs upon cooling and heating, and xray diffraction established the fcc sublattice of La, as in the LaH$_{10}$ structural prediction.
Superconductivity was confirmed by Drozdov and collaborators, initially at 215K [\onlinecite{Drozdov2018a}] but soon thereafter up to 250K.[\onlinecite{Drozdov2019}]. The latter paper reported vanishing resistivity around 170 GPa, a H isotope effect, T$_c$ decreasing with applied magnetic field, and evidence of the predicted crystal structure.  

Magnetic measurements by Eremets {\it et al.} \cite{EremetsData,Minkov2022a,Minkov2022b}
were noted in Sec. V.C.2, the latter introducing a new measurement technique
probing flux pinning  in the very small samples. For LaH$_{10}$, the
behavior was characteristic of conventional Type II behavior, with the derived
values of $\Lambda(0)_L$=14.4 nm, $\xi$=14.4 nm, and Ginzburg-Landau parameter
$\kappa_{GL}\approx$9..

\subsection{Yttrium superhydrides: the third discovery}
\vskip 2mm \noindent 
\subsubsection{Theory} 
The Y-H system has a more extensive history than S-H or La-H systems, partly because some of the design of La-H materials included the isovalent Y-H system. Li {\it et al.} reported\cite{YLi2015} materials design for this system soon after their 2014 work on SH$_2$ discussed in Sec. V.B. Their 2015 structure search at high pressure identified YH$_3$, a bct YH$_4$ lattice with both atomic H sites and H$_2$ units, and bcc  YH$_6$ with a sodalite structure, [Fig.~\ref{fig:structures}(b), Y surrounded by 24 H atoms] as promising candidates. The latter two had predicted T$_c$ around 90K and 260K, respectively, in the (encouragingly low) 120 GPa pressure range. 

The YH$_6$ prediction was startling on two counts: predicted T$_c$ was 30\% above the already remarkable SH$_3$ value of the previous year, and the required pressure was somewhat 
lower. YH$_6$ contains some very strongly coupled H modes at comparatively low frequency (not far from instability), accounting for $\lambda\approx 3$ and high T$_c$ in spite of the significantly lowered phonon energy scale $\omega_{log}$ (the logarithmic frequency moment \cite{AllDyn}). However, anharmonicity and non-linear EP coupling can change predictions, especially when there are nearly unstable modes. These stoichiometries have not been reported in experimental studies as of 2022.

Peng {\it et al.} proposed \cite{Peng-PRL2017} a focus on hydrogen clathrate structures as a route to RTS. Results for ${\cal M}$H$_n$ structures with ${\cal M}$=Y and La, for $n$=6, 9, 10, and ${\cal M}$=Sc for $n$=6 and 9 were presented. 
The results pertinent for this discussion are for two yttrium hydrides. YH$_9$ at 150 GPa has the largest calculated coupling: $\lambda$$\approx$4, T$_c$$\approx$250K.
However, YH$_9$ at 400 GPa has an even larger predicted T$_c$: T$_c$$\approx$290-300K with a smaller $\lambda$$\approx$2.3. Considering the differences in structures and in optimum pressures, even their substantial amount of data only begins to provide guidelines for just what factors are most important in promoting RTS.

In 2019 Heil {\it et al.} predicted \cite{Heil_YH9} clathrate-like structures from their structural search, focusing on YH$_6$ and YH$_{10}$. 
The results for YH$_6$ were similar to those of Li {\it et al.} \cite{YLi2015}, with calculated anharmonic effects accounting for some of the differences. T$_c$ was predicted to be similar, 275K, for YH$_6$ at 100 GPa, an encouraging result for the efforts to produce and retain high T$_c$ hydrides at more accessible pressures. 
For YH$_{10}$ predicted T$_c$=300K around 300 GPa was obtained, similar to the results of Peng {\it et al.} \cite{Peng-PRL2017}  

For this compound a remarkably large coupling $\lambda$$\approx$4.5 was reported, reproducing reasonably well results of Peng {\it et al.} \cite{Peng-PRL2017} This value is among the largest values from DFT-Eliashberg theory for a real (if still only predicted) compound.
This large value of $\lambda$ `benefits' from very soft phonons, that is, being very close to a dynamical instability, which is a typical occurrence in several crystal classes. \cite{Quan2019} 
When this occurs, corrections for anharmonicity, quantum fluctuations of H, and nonlinear EP coupling become necessary to pin down the critical pressure for instability as well as for the most complete prediction of T$_c$.  Generally but especially in hydrides, low frequency modes do not promote T$_c$ as much as their contribution to $\lambda$ would suggest \cite{Quan2019,Roadmap2021} (see Sec. VI.B).

\vskip 2mm \noindent 
\subsubsection{ Experiment} Experimental verification of the prediction of high T$_c$ in the yttrium hydride YH$_9$ was announced \cite{Kong2019} in 2019, and published in 2021: Kong {\it et al.} \cite{Kong2021}, T$_c$=243K at 200 GPa in space group $P6_3/mmc$, with the expected structure being clathrate-like.
The compound YH$_{10}$ predicted to have higher T$_c$ was not observed in their experiments, which covered certain regions of phase space up to 410 GPa and 2250K. 
Snider {\it et al.} in 2021 provided data \cite{Snider2021} indicating T$_c$ up to 262 K for a sample with superconducting phase of stoichiometry likely close to YH$_9$ based on Raman data. This maximum T$_c$ occurred around 180 GPa.  Extension and some degree of confirmation was provided when T$_c$=253K was obtained in (La,Y)H$_n$ mixtures by Semenok {\it et al.} \cite{Semenok2021}

As mentioned, other regions of the Y-H phase diagram have been predicted to display high temperature superconductivity. In 2021 Troyan {\it et al.} reported \cite{Troyan2021} T$_c$ up to 224K at 166 GPa in cubic $Im{\bar 3}m$ YH$_6$. This compound is an example of strong effects of anharmonicity due to the structure and small proton mass. The calculated values are $\lambda$=2.4 using harmonic quantities, reduced to  $\lambda=1.7$ with anharmonic corrections.  Anharmonicity considerably hardens the lower frequency H phonons, giving a calculated value of $\omega_{log}$=115 meV and T$_c$ in the range 180-230K depending on some choices. Nonlinear coupling corrections may be important to obtain the best predictions. Items to be aware of when comparing predictions with data have been noted in Sec. V.A. 

\section{Current challenges}

\subsection{Further theoretical guidance}
The progress in raising T$_c$ over eleven decades of time is illustrated in 
Fig.~\ref{fig:maxTc}. The lesson of the past was that increases in T$_c$ 
cannot be foretold. On the other hand, the discovery of new superconductors 
with T$_c$ approaching room temperature in compressed hydrides has been 
enabled by material-specific theory and computational materials design, 
after which near room temperature superconductivity was predicted, then 
verified by experiment. Intense effort continues toward discovering higher 
temperature or more accessible superconductors. Given the rapid progress, 
the way forward suggests optimism, with some understanding of the microscopic 
processes being partnered by computational power in the search. Considering 
that the full DFT-Eliashberg results can be dug into at any level of detail 
that one wishes, one can expect that helpful understanding 
will soon begin to emerge. Section VI.B indicates the current level of 
understanding and identifies the direction needed for further analysis. 

In the previous section, some of the calculated values of EP coupling strength $\lambda$ for the verified hydrides (and some others) have been mentioned. The superconducting hydride sample is small, but it has already been clear that $\lambda$ in itself is a unreliable indicator of T$_c$; increasing $\lambda$ should not be the primary goal. 
Past examples show that, within a given structure, $\lambda$ is increased
by lowering frequencies, which (often quickly) leads to lattice
instability.  Indications from Allen-Dynes analysis \cite{AllDyn} through more 
recent studies \cite{Quan2019} indicate that increasing the electron-proton 
scattering matrix element ${\cal I}_H$ should become the focus of attention, 
always assisted by large $N(0)$ of course. 

Practically nothing is understood at present about what keeps ${\cal I}^2$ large in the RTS materials when the electron gas is being compressed to higher density, and (in the simplest picture) should be screening more strongly, thereby reducing ${\cal I}^2$.  $N(0)$ itself is usually normal in size, which may help to promote stability. Several groups have noted that the H-atom DOS $N_H(0)$ per unit volume is likely to be the relevant quantity, 
but it does not seem to correlate strongly with T$_c$.

With leadership in materials design by the theoretical and computational modeling communities, and materials discovery by high pressure experimenters with increasingly advanced techniques, the six decade old hope for room temperature has essentially been achieved: T$_c$ up to the 250-260K range has been reproduced and accepted by
2020, in substantial measure  due to agreement with prediction. 
The hurdle for closer study and application is that 150-250 GPa pressure is required. This observation directly suggests two future primary  goals: (1) yet higher T$_c$ at high pressure, and ultimately superconducting atomic hydrogen, for the advancement of scientific achievement and for knowledge base, and (2) producing or retaining HTS to much lower, or preferably even ambient, pressure, for applications. The following comments provide items that could lead to advancement toward these goals.  Some progress in these areas are noted in following subsections.

\vskip 2mm \noindent
\subsection{Analysis of H coupling}
Analysis of strong EP coupling strength, especially at high frequency, is crucial in understanding how to increase T$_c$, as long as it contributes to both the frequency scale and to $\lambda$, and also helps to avoid structural instability. However, accomplishing this in compounds is more involved and less transparent than in elements because the relevant 
quantities -- $N(0)$, matrix elements, masses, and phonon frequencies -- are mixtures of the constituent atoms and their interplay. 

Hydrides are special, besides their high T$_c$, because the large difference in the metal mass and the proton mass separates the phonon spectrum $F(\omega)$ and hence the Eliashberg spectral function $\alpha^2F(\omega)$ into separate frequency ranges: \cite{Quan2019} low frequency metal acoustic modes separated by a gap from high frequency H optic modes. 
This separation of frequency regions allows the isolation of contribution
from each atom
\bea
\lambda=\lambda_{\cal M} + \lambda_H; \lambda_X
     =\frac{N_H(0){\cal I}_X^2}{M_X\omega_{2,X}^2}
\label{eqn:separatelambda}
\eea
for $X$ equal to metal ${\cal M}$ or hydrogen $H$. This procedure extends the productive
analysis of elemental superconductors to allow identification of 
the {\it individual atomic contributions}
for compressed metal hydrides. The contribution of the metal atom, discussed below,
is enlightening. The comparison of hydrogen contributions, and the individual
components contributing to $\lambda$, for the various hydrides becomes possible,
with an example being given in Fig.~\ref{fig:6panels}.

This separation is shown for the spectral function of SH$_3$ in Fig.~\ref{fig:SH3}. The separation of the degree of participation is almost complete.  Whereas it has been common to quote the separate metal and H contributions to $\lambda$, as quoted for a few examples above, the capability of separating all of the atomic information is available from $\alpha^2F$.  This deeper analysis is important because $\lambda$ alone is a poor indicator of T$_c$. What is of prime interest is the contribution to T$_c$ (versus $\lambda$, energy gap, or other properties) from each atom. 

\begin{figure}[tbp]
  \hskip -4mm
  \includegraphics[width=1.05\columnwidth]{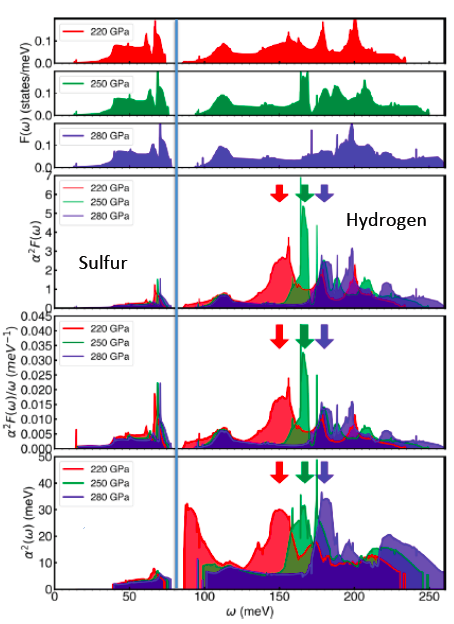} 
 \vskip -6mm
  \caption{Color-coded plots of several spectral functions of
  SH$_3$ at the three pressures indicated in the top  three
  panels,
  all above the optimum pressure of 
  160K. The main point is to illustrate the separation of S and 
  H modes (the solid vertical line at 80 meV), allowing analysis 
separately of their coupling strengths 
  and contribution to T$_c$. The blue (solid) vertical line marks this 
  separation, lying in the gap around 80 meV between S (left, 
  low frequency) and H (right, high frequency) vibrational modes. 
 Top three panels: phonon density of states $F(\omega)$ for each 
  of the three pressures denoted. 
 The main effect of pressure is to push some H modes to higher 
  frequency. The bottom three panels show, respectively, 
  $\alpha^2F(\omega)$, $\alpha^2F(\omega)/\omega$, and the ratio 
  $\alpha^2(\omega)$=$\alpha^F/F$ that indicates the mean coupling 
  matrix elements' strength at frequency $\omega$. The arrows 
  indicates peaks of interest. 
 The low frequency region is subject to numerical noise and has 
  little weight so it has been cut out. Taken  
  from Quan {\it et al.} 2019.}
  \label{fig:SH3}
\end{figure}

Analysis based on the frequency separation of metal and H modes was 
reported by Quan {\it et al.} \cite{Quan2019} for five binary hydrides 
from three crystal structure classes. A phase diagram illustrating some 
aspects of the analysis is shown in Fig.~\ref{fig:structures}(d). 
A primary indication is of rough phase boundary identifying an 
``island of instability'' in the high T$_c$, but
lower $\omega_2$ regime. Within a high T$_c$ phase, this instability
is encountered as pressure is {\it lowered}, with $\lambda$ and T$_c$
increasing until instability is reached. 
For brevity, only a few other results of this trend study
will be discussed here.

The first observation in this study is startling. Although it is an 
appreciable fraction of $\lambda$, the {\it metal atom contribution} to 
$\alpha^2F$ {\it affects T$_c$ very little}, and {\it sometimes lowers it}. 
This occurs because coupling to low frequency vibrations lowers 
the phonon frequency scale -- the prefactor in T$_c$ in 
Eqs.~\ref{eqn:BCS}, \ref{eqn:McM}, and \ref{eqn:AD} -- off-setting the 
increase in $\lambda$. 
Quan {\it et al.} explained why this does not violate the Bergmann-Rainer 
`theorem': additional coupling at one frequency affects the phonon 
spectrum and coupling at other frequencies. This conclusion can be confusing,
because the metal contribution to $\lambda$ is evident and often emphasized,
while the effect on the frequency scale is never calculated, therefore remaining
invisible. 
This result focuses attention for analyzing and understanding T$_c$ on 
hydrogen alone
-- neglect the metal -- thereby providing a focus for a more transparent 
understanding of T$_c$ in terms of hydrogen properties.

\begin{figure*}[tbp]
  \hskip -4mm
  \includegraphics[width=1.95\columnwidth]{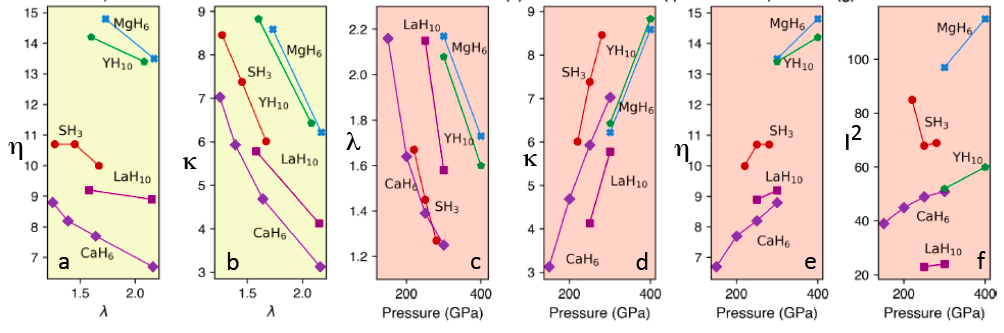} 
 \vskip -2mm
  \caption{Atomic H quantities in the indicated superconducting compressed hydrides, obtained by taking advantage of the spectral separation of metal acoustic modes and H optic modes.
  Corresponding calculated values of T$_c$ can be seen in Fig.~\ref{fig:structures}(d).
  (a,b) $(\eta,\lambda)$ and $(\kappa,\lambda)$ phase planes, respectively, indicating small (resp. large) relative variations.
  (c-f) Pressure variations of the indicated quantities, showing the magnitudes and rates of variation with pressure.
  Units: $\lambda$, unitless; $\eta=N_{\uparrow}(0){\cal I}^2$ and 
    $\kappa=M_H\omega_2^2$, eV/\AA$^2$; ${\cal I}^2$, eV$^2$/\AA$^2$.
  The background color (or different shadings in grayscale figures) indicate
  the different types of plots, see the axis labels.
  Adapted with permission from Quan {\it et al.} 2019.
  }
  \label{fig:6panels}
\end{figure*}

Second, and following from the first: $N_H(0)$ and $\omega_{2,H}^2$ are 
calculated and the mass $M_H$ is fixed. This makes it possible to extract 
and analyze the least understood and very important quantity: the matrix 
element ${\cal I}_H^2$ factor in Eq.~\ref{eqn:separatelambda} for scattering 
electrons from the moving H atom. Quan {\it et al.} provide a discussion 
of this and several other aspects of EP coupling in their five chosen
compounds, arguing that understanding of ${\cal I}_H^2$ is the main missing 
link in the understanding of EP coupling in compressed hydrides.
A synopsis of some of their analysis is shown in Fig.~\ref{fig:6panels}. 
In the ($\eta_H,\lambda_H)$ phase plane of Fig.~\ref{fig:6panels}(a) 
and the pressure dependencies of $\eta_H, I_H^2$ in 
Fig.~\ref{fig:6panels}(e), (f), respectively, the example hydrides seem 
to self-organize into groups. In the ($\kappa_H,\lambda_H)$ plane and the 
other two pressure dependencies, the magnitudes and pressure behaviors 
of the five phases show 
considerable overlap but clear trends.  Note that ${\cal I}^2$ tends to increase 
with pressure, but (for example) the magnitudes at 300 GPa differ by a 
factor of 3.
          
As mentioned earlier, the strong tendency for ${\cal I}^2$ to increase with 
pressure is counterintuitive, since the increased electronic density 
under volume reduction would seem to provide increased screening of 
H motion. This should reduce the ${\cal I}_H^2$ factor given 
in Eq.~\ref{eqn:matrixelement}.
The change in potential due to H displacement 
can be evaluated within linear response theory, either from the
dielectric constant with all local field terms (highly intensive
numerically) or numerically from linear response, which is the method 
of choice and is still computationally intensive enough to sometimes limit
$Q$ (phonon) point grids.

The pressure increase in ${\cal I}_H^2$ indicates there is more physics to be 
understood. One can compare with a simple and efficient approximation from 
Gaspari-Gy\"orffy theory, \cite{GG} which uses a rigid atomic potential 
displacement model. The result requires negligible computation, but 
involves phase shifts of the potential that provide only a limited physical
understanding. The Gaspari-Gy\"orffy model was applied to SH$_3$ by 
Papaconstantopoulos {\it et al.} \cite{Papa2015}, finding indeed that 
$\eta_H^{GG}$ also rises steeply and nearly linearly from 
18 eV/\AA$^2$ at 200 GPa to 25 eV/\AA$^2$ at 300 GPa. 
The underlying mechanisms remain unclear.

Fig.~\ref{fig:6panels}(f) indicates the extracted value 
(without approximation) of ${\cal I}_H^2$ is 10-11 eV$^2$/\AA$^2$ 
around 200 GPa, indicating a 70\% overestimate by Gaspari-Gy\"orffy 
theory, with its neglect of screening. The calculations of 
Quan {\it et al.} were not extended to higher pressure, so a more 
complete comparison is not available.  Hutcheon {\it et al.} have 
used Gaspari-Gy\"orffy theory to give a quick estimate of $\eta$ 
in their machine learning study. \cite{Hutcheon2020}.

\subsection{How to produce strong H coupling}
\subsubsection{The metal-induced atomic hydrogen paradigm}

The leading paradigm in compressed hydrides, from Ashcroft, is the need 
to break, or deter, the strong H$_2$ molecular bond or other molecular bonds 
(viz., CH$_4$, NH$_3$), which will move the bonding band below E$_F$ and 
push the antibonding bands above E$_F$, leaving little or no H contribution 
at the Fermi level. Breaking up molecules, leaving open $1s$ shell atomic 
hydrogen rather than molecular hydrogen, is the paradigm followed by the
three first discoveries, and by nearly all of the predicted
hydrides with T$_c$>100 K. Being the leading concept for 60 years, 
little more needs to be said about this paradigm,
but this being a new field of investigation and discovery, it is important
to look for other paradigms.  

\vskip 2mm \noindent
\subsubsection{An alternative paradigm: activating bonding states}
The RTS examples introduced in Sec. V are from the anticipated class, in which hydrogen becomes atomic (no overt covalent bonding) and predominantly H $1s$ bands lie at the Fermi level and are primarily, almost overwhelmingly, responsible for high T$_c$. Computational explorations of ternary hydrides (see below) have found that at too low pressures, cells with molecular H$_2$ or H-rich molecules provide the stable structures. In such compounds, bonding and antibonding H $1s$ levels are split away from the Fermi level (below and above, respectively) and H vibrations provide little or no EP coupling. It seems that a guiding principle is that atomic H dominance leads over other productive possibilities, and its coupling strength requires further attention. Yet another possibility has arisen. 

{\it Li$_2$MgH$_{16}$}. 
Sun {\it et al.} reported calculation \cite{ysun2019} of T$_c$ around 475K 
at 250 GPa for Li$_2$MgH$_{16}$, 
which is best pictured as a lattice of MgH$_{16}$ clathrate-like units intercalated by Li.
The large unit is rather stable, while the Li donor adds electrons, lends the H sites
a more metallic character, and doubles the value of N$_{\uparrow}$(0). The maximum H frequency was
2400 cm$^{-1}$, in the same range as other HTS hydrides at 
similar or somewhat lower pressure. As in other
hydrides, T$_c$ decreases with increasing pressure while 
the frequency spectrum 
increases.

The distinction that makes T$_c$ higher than in other compressed 
hydrides is unclear. The small
mass of Li results in the overlap of its phonon projected 
density of states with that of H; Mg is lower and nearly separate.
For $\lambda$=3.35 in this P-T$_c$ regime, roughly 1.75 can be
ascribed to the low frequency metal atoms ($\omega<20$ THz), with 
1.60 arising from H modes
extending up to 70 THz. The two H sites contribute very differently 
to the bands crossing
the Fermi level (Y. Quan and W. E. Pickett, unpublished). 
Comparative analysis with 
other HTS compressed hydrides is needed to
obtain insight into the origins of high T$_c$.

{\it LiB$_2$H$_8$.} As another example, 
Gao {\it et al.} in 2021 reported a designed (predicted) H-rich 
system \cite{MGao2021} in which high T$_c$ (though not RTS) arises 
in a distinctive manner. 
The material is one in which BH$_4$ units (identifiable molecules) 
lie in interstitial positions within a bcc potassium sublattice, 
comprising KB$_2$H$_8$ = K(BH$_4$)$_2$, with identifiable BH$_4$
molecules. 
Extrapolating from results on other ternaries, such a compound having only 
molecular hydrogen should be unpromising. 
However, the chemistry (more specifically, the Madelung potential) is 
such that each molecular BH$_4$ radical (likely unstable in itself, 
lacking the extra electron that stabilizes methane CH$_4$) obtains 
$\frac{1}{2}$ electron from the K ion, leaving the uppermost (least 
strongly bound) molecular orbital half-empty. The resulting radicals 
are stable within the sublattice of positive K ions, and the compound 
is predicted to be dynamically stable.  

KB$_2$H$_8$ is calculated to be metallic, but with the character 
of a heavily hole-doped wide-gap insulator. 
This leaves covalently bonded bands that are strongly coupled to B-H bond-stretch modes at the Fermi level, a close analog \cite{An2001} of MgB$_2$ with its T$_c$=40K. 
The calculation of Gao {\it et al.} gave T$_c$$\approx$140 K at the modest
pressure of only 12 GPa, arising from very large $\lambda$=3 but an 
unusually low frequency scale $\omega_{log}$=33 meV (100+ meV is 
more typical of RTS hydrides, but at pressures of 150 GPa and higher). 
This is a three dimensional extension of the argument that such MgB$_2$-like systems can be optimized to produce much higher EP-coupled superconductivity. \cite{WEP2008}
Further improvements in this direction seem possible. 

However, with such a large hole density that covalent bonds may 
be unstable, LiB$_2$H$_8$ may not be a thermodynamically stable 
composition. This scenario played out in Li$_{1-x}$BC, where for 
$x\sim 0.2-0.3$, T$_c$ up to 75 K was inferred. \cite{Rosner} 
Substantial experimental effort \cite{Fogg} could not produce the 
desired structure at the target doping levels, obtaining instead 
distorted and disordered materials. However, Sr$^{2+}$  doping on 
the K$^{1+}$ sites, lowering the hole doping level, and broader 
synthesis routes may provide pathways to desired materials. \cite{Nakamori}

\subsection{Increasing accessibility; metastable structures}

\subsubsection{Lowering the required pressure}
Of growing concern is to find, perhaps by design and discovery or 
perhaps by serendipity, materials that will retain their high pressure
high values of T$_c$ to lower pressures, with the intention of finding 
applications.  Most of the binary hydrides have been explored with 
computational means. \cite{zurek2018a,FloresPerspective2020} Analysis of 
the results remains to be done, and unfortunately published results often do 
not provide much of the information that is required, including the 
atom-specific quantities in $\lambda$ in Eq.~\ref{eqn:lambda}. 
The study of Quan {\it et al.} has initiated such analysis, \cite{Quan2019} 
but was limited to a few binary hydrides for which their recalculations 
provided the data required for the analysis of the electronic properties.  

Separately but equally valuable is an improved understanding of the stability, 
or lack thereof, of high T$_c$ materials composition and structures. Considering 
broadly, there are several examples of this scenario: a high T$_c$ 
material is discovered (either computationally or experimentally) and 
its structure understood; pressure is lowered and T$_c$ (and calculated 
$\lambda$) increases but a phonon branch is lowered; a structural phase 
transition occurs at a critical pressure $P_{cr}$; in the low pressure 
phase T$_c$ is much lower or perhaps vanishing; the structure of the new 
phase includes H-rich molecular units, including possibly H$_2$, without 
much or any atomic hydrogen. One question being addressed is: how can 
this process be pushed to lower $P_{cr}$, or even (ideally) to ambient pressure.    


\vskip 2mm \noindent 
\subsubsection{More complex hydrides; speeding searches}
After the design and discovery reported in the sections above, emphasis has broadened. Higher values of T$_c$ are of course still of great interest; after all, applications at room temperature will require T$_c$ around 30\% higher (375-400K), or
even~higher for high current density applications. Given the considerable number of binary hydrides that have been modeled and mined for high T$_c$ (here meaning roughly, T$_c$$>$100K), useful for applications), searches are being extended to ternary hydrides. 

The palette of ternary hydrides is much larger than that of binaries, 
thereby opening new candidates and new computational challenges. 
So far the emphasis has been on the more H-rich possibilities, viz. A$_i$B$_j$H$_n$ with small $i$ and $j$ and larger $n$. With atoms A and B selected from the (say) 60 most reasonable choices of elements, and with concentration $n$ ranging up to 12, this class has of the order of $10^4$-$10^5$ formulae, and for each of these, many crystal structures are possible. Given this complexity, techniques in high-throughput computing, data set construction, data mining, coupled with machine learning, are being applied to the search for promising candidates, but a full search is not in sight. Background on these activities can be found in the {\it 2021 Roadmap} compilation. \cite{Roadmap2021}
There are too many reports already on ternary hydrides with too little analysis to attempt to provide an overview. Several candidates have predicted T$_c>$100K, however explorations of the generalized convex hull to find the most stable stoichiometries have been limited.

\subsubsection{Exploring higher pressures} 
While discussion of advancing high-pressure techniques is well beyond the scope of this article, it should be mentioned that experimental extensions to achieve higher pressures more readily, and to adapt measurement techniques to obtain more general data on the samples, are continually pursued in the high pressure laboratories that have contributed to high T$_c$ hydrides at high pressure.


 \section{Regularities in Compressed Hydrides} 
While this article is not intended as a topical review of compressed hydrides, it 
should have raised questions, and addressed some of them,  about the properties 
that provide close approach to room temperature superconductivity. 
Several features that appear to be important clues, {\it i.e.} to have some generality, 
have been identified.
\vskip 2mm \noindent $\bullet$ A fundamental question is: what structure types
of compressed hydrides are favored, and in what pressure ranges? Some guidelines
seem to have arisen. In the lower pressure range, hydrogen molecule phases arise
and are poor superconductors or, often not reported, insulators. At higher pressure
(say, 150-300 GPa, atomic hydrogen structures are frequently favored, with H-caged
metal structures ({\it e.g.} clathrate) being common. At still higher pressures
(with fewer published examples) less common structures are predicted; for example,
one has several layers of hydrogen followed by a few layers of metal, suggestive of
incipient phase separation. This area is a complex one, and the reviews and overviews
mentioned below should be consulted.

\vskip 2mm \noindent 
$\bullet$ An overriding question -- what properties enable room temperature 
superconductivity -- remains open to clarification. 
It is not large $\lambda$ {\it per se}; $\lambda\approx$2-3 is similar to that 
in Pb-Th-Bi alloys, with their T$_c$$\approx$10K. These low T$_c$
materials have very soft phonons,
due to heavy masses and nondescript $s$-$p$ bonding. 
Compressed hydrides have frequency scales up to 150-200 meV 
(1750K-2300K), compared to those
heavy atoms with frequency scales of  50-75K. The enabling feature 
is, as Ashcroft foretold, 
producing H-derived modes at high frequency while retaining strong coupling 
to Fermi surface carriers. 
This is, unfortunately, more an observation than an answer.
\vskip 2mm \noindent $\bullet$ Strong coupling. Barring innovations, 
larger  $\lambda$ 
should not be the overriding aspiration. There are numerous cases, including compressed 
hydrides (mentioned above), where increasing $\lambda$ increases T$_c$ but rapidly 
encounters soft phonons and lattice instability. $\lambda$ 
is an unreliable  descriptor for 
a high T$_c$ search. Clue: $\eta/M$, independent of frequencies, is much better. \cite{AllDyn,Quan2019} 
Since higher frequency scales will likely require ever higher pressures, the 
alternative seems to be pushing strong coupling to the high frequency region. 
In compressed hydrides `bond stretch' modes seems not to be a dominant consideration.
Generally H-H `bonding' is not a clear feature; H-metal bonding is more often 
a topic of consideration. 
\vskip 2mm \noindent $\bullet$ Various groups \cite{semenok-distribution,Belli-Errea2021} 
have observed that atoms in columns I, II, and III provide, with occasional exceptions, 
the binary hydrides with high T$_c$. These atoms have low electronegativity, readily 
donating electrons to the H sublattice(s). The resulting negatively charged H atoms 
(versus neutral, half filled entities such as the H$_2$ molecule) promote breaking of 
H bonds and producing metallic ground states. This factor might also be related to the 
retention of strong coupling to high energy vibrations.
\vskip 2mm \noindent $\bullet$ Is a higher concentration of H the key?
The indications are that a large fraction of H states at the Fermi surface, {\it i.e.} 
reflected in a large ratio $N_H(0)/N(0)$, is not a clear 
determining factor, or at least is not essential. 
This is a straightforward band structure quantity, and cannot be estimated 
before the band calculation is done, because band structure effects cause 
structure in $N(\varepsilon )$. 
An example for this item: SH$_3$ is somewhat exceptional, with a strong van Hove peak 
at the Fermi level. \cite{Ghosh2019}, \cite{Pickard2019}, 
\cite{FloresPerspective2020} have highlighted the 
$N_{\uparrow}(0)$ factor in compressed hydrides.
\vskip 2mm \noindent $\bullet$ Producing ``atomic hydrogen,'' as opposed to 
molecular-bonded hydrogen, has been the overriding objective, and so far
the productive one. Pressure will 
eventually decompose hydrogen-rich molecules, but other methods (viz. doping) should 
be kept in mind. Doping holes into bonding states may be promising, \cite{MGao2021} 
but too much doping will make a stable phase prone to structural instabilities.
\vskip 2mm \noindent $\bullet$ Lattice instabilities. In several cases of high 
T$_c$ hydrides, it has been found that within a given phase T$_c$ decreases 
with increasing pressure. Conversely, T$_c$ increases as pressure is decreased, 
$\lambda$ increases, and the modes giving the increase in $\lambda$ are renormalized 
to lower frequency and then become dynamically unstable. For the five systems 
illustrating this self-limiting process in Fig.~\ref{fig:structures}(d), there 
is an indication of a phase boundary for binary hydrides in the $\omega_2$-T$_c$ plane. 
\vskip 2mm \noindent $\bullet$ Naturally, the lattice stiffness 
$\kappa_H=M_H\omega_{2,H}^2$ 
increases with pressure. However, strong coupling is far from uniform throughout 
the H-derived optic modes. Pressure does increase the frequency scale, but  
appearing squared in the denominator it decreases $\lambda$. 
This trade-off has long been a persistent issue when 
pursuing higher T$_c$ superconductivity. More focus needs to be aimed at
increasing $\eta_H$.
\vskip 2mm \noindent $\bullet$ The scattering efficacy of the vibrating H atom, 
${\cal I}_H^2$, increases with pressure, according to current information. \cite{Quan2019,Papa2015} 
The origin of this simple fact is unclear, but the theoretical and computational 
means to understand it is available within DFT codes (but requiring proper extraction
and analysis). 
\vskip 2mm \noindent $\bullet$ For the small set of examples that has been studied,  
${\cal I}`_H^2$ varies from one to another over the pressure range of interest, by a factor 
of three, and $\eta$ by a factor of two, \cite{Quan2019} see Fig.~\ref{fig:6panels}. 
This difference can be attributed partially to broadening of the (largely H) 
bandwidth with
increasing pressure, hence tending to decrease $N_H(0)$.
Again, the origin is unclear but the detailed computational theory exists to analyze 
this fact in detail.

\vskip 2mm
A careful study of the ``regularities'' listed above will reveal 
repetition and apparent 
inconsistencies and contradictions. Example: strong coupling at high frequency is 
what is really important, but also $\eta=N_{\uparrow}(0){\cal I}^2$ 
(which is independent of frequency) 
is what really matters at strong coupling. Such various viewpoints are what must 
be confronted in the quest for higher T$_c$ at lower pressure. 
Also, nearly universally high T$_c$ has been couched separately 
in terms of $\lambda$
and one characteristic phonon frequency, viz. $\omega_{log}$.
This approach my be misguided ({\it i.e.} not the most profitable) for progress.
It has been noted that, for the five compressed hydrides mentioned in
previous sections. The simple relation T$_c$$\propto$A, where A is the
area under $\alpha^2F$, works quite well in spite of its simplicity.
 \cite{Quan2019} Improvements in the T$_c$ equation and in understanding
by a generalization to another 1-3 more characteristics obtained from
$\alpha^2F$ might be fruitful. 

\section{Epilogue} 
A number of overviews \cite{Pickett-PT,Pickard2019,Boeri-Road,Shimizu2020} and more 
extensive collections \cite{zurek2018a,zurek2018b,FloresPerspective2020,Roadmap2021,Zurek2022b,semenok-distribution,zurek2022} 
of predictions of hydride superconductivity are available. The achievement 
of (near) room temperature superconductivity has 
stimulated extension of high pressure techniques and the enabling of additional 
measurements, in step with improved analysis and interpretation of 
data. \cite{Hemley2019,Guan2021}  

The first point of this perspective was provided in Secs. II and III, which summarizes 
the sequence of theoretical and algorithmic advances, 
followed by numerical implementation, 
that have produced an accurate, material-specific theory of 
EP superconducting T$_c$ as well 
as several superconducting properties not discussed here, mostly stemming from the 
complex superconducting gap $\Delta(\omega,T)$. The three initial advances predicted by 
the theory and then confirmed by experiment are discussed in Sec. V:\\
$\bullet$ SH$_3$, 200K at 100 GPa \\
$\bullet$ LaH$_{10}$, 260K at 200 GPa\\ 
$\bullet$ YH$_9$, 240-260K around 250 GPa. \\

This article is intended to provide an overview of the theory-driven forces behind 
the design and discovery of room temperature superconductivity. The experimental
effort on hydrides has been impressive as well.  Room temperature 
superconductivity was a much discussed but distant goal in the 1970s, but expectations 
faded after 13 years with no increase in T$_c$. The discovery of high T$_c$ cuprates 
revived the dream to some extent, but the focus of research soon reverted to an 
intense study of the properties 
and mechanism(s) (versus magnitude of T$_c$) of superconducting quantum materials,
a topic that remains a leading paradigm of condensed matter physics that 
is being broadened to other classes, properties, and applications. 
In terms of temperature, HTS has 
been superseded only by compressed hydrides -- 
the long-sought room temperature superconductors.

\section{Acknowledgments}
This overview has benefited from discussions with numerous colleagues over the past few 
years. This perspective was stimulated by communications with R. E. Cohen and R. J. Hemley, 
who confirmed my view that the impact of theory and materials design on searches for 
room temperature superconductivity should be chronicled. A conversation with M. R. Beasley
stimulated the investigation of order parameter phase fluctuations at room temperature.
I acknowledge Yundi Quan and Soham Ghosh for collaboration on this topic in recent years,
and F. Giustino for important help on typesetting for this journal.

%

\end{document}